\documentclass{emulateapj}

\def\gsim{\mathrel{\rlap{\lower 4pt \hbox{\hskip 1pt $\sim$}}\raise 1pt
\hbox {$>$}}}
\def\lsim{\mathrel{\rlap{\lower 4pt \hbox{\hskip 1pt $\sim$}}\raise 1pt
\hbox {$<$}}}

\usepackage{color}

\begin{document}

\title{Type Ia Supernovae in the First Few Days: \\
Signatures of Helium Detonation versus Interaction
}  

\author{Keiichi Maeda\altaffilmark{1}, Ji-an Jiang\altaffilmark{2,3}, Toshikazu Shigeyama\altaffilmark{4}, Mamoru Doi\altaffilmark{2,4}}

\altaffiltext{1}{Department of Astronomy, Kyoto University, Kitashirakawa-Oiwake-cho, Sakyo-ku, Kyoto 606-8502, Japan; keiichi.maeda@kusastro.kyoto-u.ac.jp .}
\altaffiltext{2}{Institute of Astronomy, Graduate School of Science, The University of Tokyo, 2-21-1 Osawa, Mitaka, Tokyo 181-0015, Japan}
\altaffiltext{3}{Department of Astronomy, Graduate School of Science, The University of Tokyo, 7-3-1 Hongo, Bunkyo-ku, Tokyo 113-0033, Japan}
\altaffiltext{4}{Research Center for the Early Universe, Graduate School of Science, The University of Tokyo, 7-3-1 Hongo, Bunkyo-ku, Tokyo 113-0033, Japan}

\begin{abstract}
The mechanism for the early-phase blue and excessive emission within the first few days, reported so far for a few type Ia supernovae (SNe Ia), has been suggested to be interaction of the SN ejecta either with a non-degenerate companion star or circumstellar media (CSM). Recently, another mechanism has been suggested within the context of the He detonation-triggered SN scenario (i.e., double detonation scenario or He-ignited violent merger), in which the radioactive isotopes in the outermost layer of the SN ejecta produce the early emission. In this paper, we investigate properties of the early-phase excessive emission predicted by these different scenarios. The He detonation scenario shows different behaviors in the early flash than the companion/CSM interaction scenarios. Especially clear diagnostics is provided once the behaviors in the {\em UV} and in the optical are combined. The spectra synthesized for the He detonation scenario are characterized by the absorptions due to the He detonation products, which especially develop in the decay phase. We further expect a relation between the properties of the early-phase flash and those of the maximum SN emission, in a way the brighter and slower initial flash is accompanied by a more substantial effect of the additional absorptions (and reddening). This relation, however, should be considered together with the maximum luminosity of the SN, since the larger luminosity suppresses the effect of the additional absorption. With these expected features, we address the possible origins of the observed excessive early-phase emission for a few SNe.  
\end{abstract}


\section{Introduction}

Recent rapid development of transient surveys has opened up a new window of studying explosive transients within the first few days since the explosion. This window should bring us new information on unresolved problems of the stellar evolution and explosion mechanisms. Detection of a shock breakout signal from core-collapse supernovae (CCSNe) has long been awaited for decades \citep[e.g.,][]{tominaga2011, suzuki2016}, the detection of which in the optical has been claimed by \citet{garnavich2016} \citep[but see][]{rubin2017} or more recently by \citet{bersten2018}. The post shock breakout `cooling' emission has been detected for a number of CCSNe, which provides powerful diagnostics to measure the radius of the progenitor star \citep[e.g.,][]{arnett1980,bersten2012}. The early emission might also be used to clarify the nature of circumstellar matter (CSM) in the close vicinity of the progenitor, which reflects the mass loss process in the final years to decades \citep[e.g.,][]{galyam2014,morozova2017}. 

Type Ia supernovae (SNe Ia) are widely believed to be a thermonuclear explosion of a white dwarf (WD). For the small radius of the progenitor WD, the (usual) post shock breakout cooling emission is substantially weaker than in CCSNe \citep{rabinak2012}, therefore using the early emission in the first few days to measure the progenitor radius is difficult for SNe Ia. For example, \citet{nugent2011} constrained the radius of the progenitor to be $\lsim 0.1 R_{\odot}$ for the extremely nearby SN Ia 2011fe in M101 \citep[see also][]{bloom2012}, which was discovered within a day since the explosion. 

For SNe Ia, the early emission in the first few days\footnote{In this paper, we refer the first few days as the `early phase', while the photospheric phase covering the maximum light ($\sim 10 - 30$ days since the explosion) is called the `maximum phase' (note that this phase is also usually called the early phase).} has been suggested to be a powerful probe to the progenitor system. The progenitor evolution leading to the thermonuclear ignition of a WD has been actively debated over several decades. There are two popular scenarios; an accreting WD from a non-degenerate companion star \citep[single degenerate scenario or SD, e.g.,][]{whelan1973,nomoto1982,hachisu1999} and merging binary WDs \citep[double degenerate scenario or DD, e.g., ][]{iben1984,webbink1984}, among others \citep[e.g.,][]{sparks1974,soker2015}. There have been various methods proposed to tackle to this issue, either for individual cases or for statistic study. These studies suggest that there could indeed be multiple pathways toward SN Ia explosions, and observational diversities of SNe Ia may reflect diverse nature of the progenitor systems \citep[see][for a review]{maedaterada2016}. 

The diagnostics provided by the `early phase' emission has been suggested to be one of the most direct methods to address the issue. In the SD scenario, it is expected that there is a non-degenerate companion (donor) star, either a main-sequence (MS) / red-giant (RG) star, or even a He star, at the time of the explosion. The separation of the binary should be close, at an order of the size of the companion star, as the companion star must have filled the Roche lobe. The solid angle covered by the companion star as viewed from the SN ejecta is therefore large. In this case, the crush between the SN ejecta and the companion should take place within at most the first few hours. The hydrodynamical interaction creates heat and thermal energy \citep{marietta2000}. While the thermal energy is quickly lost by the adiabatic expansion, bright and blue emission is predicted to follow \citep{kasen2010}. This early phase emission may overwhelm the radiation from the main part of the ejecta powered by decay of $^{56}$Ni, depending on the size of the companion star and the observed wavelength. This idea has driven the high cadence surveys/observations of SNe Ia, and an increasing number of SNe Ia are being discovered and followed from the first few days after the explosion. Initial attempt to catch the signal resulted generally in non-detection, disfavoring the existence of non-degenerate companion stars, especially RGs \citep[e.g.,][]{hyden2010}, which should have been detected according to the prediction by \citet{kasen2010} \citep[but see][]{kutsuna2015}. However, there are now at least a few examples which are reported to show possible excessive emission in the first few days \citep{cao2015,marion2016,hoss2017,jiang2017,miller2017}, indicating that the early emission may show diverse natures (see \S 7 for details). 

As an alternative scenario, it has been suggested that the hydrodynamical interaction between the SN ejecta and dense CSM in the close vicinity of the progenitor system might also produce similar early emission \citep{piro2016}. This might naturally be a case for the SD scenario, but the DD scenario would also be accompanied by a dense `CSM' created by the binary WD merger, depending on the timing between the merger and explosion \citep{raskin2013,shen2013,tanikawa2015}. 

Finally, another mechanism which leads to the early-phase excessive emission has been independently proposed by \citet{jiang2017} and \citet{noebauer2017}. They considered a specific explosion model triggered by the detonation of He on the surface of the progenitor WD. This mode of the explosion could potentially be realized both in the SD and DD scenarios; in the former the He layer is produced by the accretion from a non-degenerate He star \citep[the double detonation scenario;][]{fink2010,woosley2011}, and in the latter the He accretion is a result of a merger of a C+O WD with either an He WD or a C+O WD with the He layer \citep[the He-ignited violent merger scenario;][]{guillochon2010,pakmor2013}\footnote{Sometimes the double detonation scenario and the violent merger scenario are distinguished from the (classical) SD and DD scenarios, respectively. In this paper, we categorize them in the SD and DD scenarios.}. The scenario is suggested to produce an SN, i.e., complete disruption of a WD, if the He detonation is sufficiently strong to trigger the thermonuclear runaway near the center of the C+O WD. \citet{jiang2017} and \citet{noebauer2017} pointed out that radioactive species synthesized by the He detonation can power the first light from SNe Ia in the first few days. \citet{jiang2017} further presented their observational data for the first robust candidate of the He-triggered SN Ia (named MUSSES1604D), discussing not only the photometric properties in the first few days, but also the spectroscopic properties around maximum light. 

In this paper, we present synthetic observables based on the He detonation scenario. The paper aims at deepening and expanding the model calculations/predictions presented by \citet{jiang2017}. Also, we aim at providing possible diagnosing observables which could distinguish the different scenarios to create the early excessive emission. In \S 2, we first describe basic ideas in each model. We first construct simple models to calculate the main features for the CSM interaction scenario (\S 2.1) and the SN-companion interaction scenario (\S 2.2). The basic ideas characterizing the He detonation scenario are described in \S 2.3, where details of the numerical simulation methods to synthesize observables for the He detonation scenario are also described. In \S 3, we present synthesized photometric properties in the first few days for each model. Characteristic photometric properties are discussed in \S 4, to clarify different characteristics expected from different scenarios. \S 5 and \S 6 focus on the He detonation scenario, where we present further predictions on the spectral features and the long term behavior extending from the early (\S 5) to the maximum phase (\S 6). In \S 7, we discuss possible mechanisms for the early excessive emissions reported for some SNe Ia. The paper is closed in \S 8 with conclusions. 

\section{Mechanisms for the Very Early-Phase Emission}

In this section, we describe the physical processes which underlie properties of the early phase emissions for the SN-CSM interaction (\S 2.1), the SN-companion interaction (\S 2.2), and the He detonation (\S 2.3) scenarios. While the CSM interaction scenario was investigated numerically by \citet{piro2016} \citep[see also] []{noebauer2016}, description of the basic process has been missing. With the description of the underlying physical processes, we construct a simple model to simulate the expected photometric properties from the CSM interaction model. The SN-companion interaction model was examined by \citet{kasen2010} and \citet{kutsuna2015}. Especially, \citet{kasen2010} not only performed detailed radiation transfer simulations\footnote{We note that the model by \citet{kasen2010} assumes the initial conditions obtained through the analytic estimate, unlike \citet{kutsuna2015} who performed (direct) radiation hydrodynamic simulations.} but presented an analytic model. In this paper, we formalize the expected photometric properties from a different viewpoint, with an analogy to the CSM interaction model. Finally, the description on the He detonation scenario is an extension and more detailed description of what was presented in \citet{jiang2017}. One of the model simulations \citep[which was initially reported in][]{jiang2017} is indeed very similar to what was presented by \citet{noebauer2017}, but the present work covers much larger model  parameter space and a range of related observables (e.g., long term evolution and spectroscopic features). 

In this paper, we assume that the density structure of the SN ejecta (without the interaction with external materials) is described by an exponential function in velocity space: 
\begin{equation}
\rho(v,t) = \frac{M_{\rm ej}}{8 \pi v_{\rm e}^3 t^{3}} \exp\left(-\frac{v}{v_{\rm e}}\right) \ , 
\end{equation}
where $M_{\rm ej}$ is the whole ejecta mass (i.e., the mass of the WD, which includes the He-layer in case of the He detonation scenario). The velocity scale is set by the kinetic energy ($E_{\rm ej}$) as $v_{\rm e} = \sqrt{E_{\rm ej}/6 M_{\rm ej}}$. This density structure mimics typical SN Ia models such as W7 \citep{nomoto1982} and frequently adopted in studying the radiation output from SNe Ia \citep[e.g.,][]{kasen2006}.

\subsection{Circumstellar Interaction}

We consider a situation in which the SN ejecta crush into the CSM with finite spatial extension ($R_0$), and all the CSM is swept up well before a significant amount of the radiation start leaking out of the system. The CSM after the interaction will form a (relatively) dense shell with a width of $\Delta R_{\rm sh}$. After the shell travels to the distance ($R_{\rm sh}$) a few times the initial extent, the hydrodynamical configuration should be frozen. After that, $\Delta R_{\rm sh}/R_{\rm sh}$ would not evolve any more, and subsequent density structure should behave in a self-similar manner. Since the pressure in the shocked material will be dominated by radiation during the initial phase when the kinematic structure is reconstructed, the shell width will be roughly determined by the compression factor of $7$, therefore $\Delta R_{\rm sh} / R_{\rm sh} \sim 0.15$ is a reasonable estimate. The shell position ($R_{\rm sh}$) will freely increase as $R_{\rm sh} \propto t$ (where $t$ is the time after the explosion, neglecting the time elapsed before the interaction). The density of the shell ($\rho_{\rm sh}$) will decrease simply as $\rho_{\rm sh} \propto R_{\rm sh}^{-3} = (V_{\rm sh} t)^{-3}$, where $V_{\rm sh}$ is a characteristic velocity of the shell. 

During the interaction, the reverse shock will penetrate down in the ejecta until the ejecta mass comparable to the mass of the CSM ($M_{\rm CS}$) is swept up \citep{chevalier1982}. This process provides a characteristic velocity of the resulting shell. Adopting the exponential profile for the pre-interaction (unperturbed) ejecta structure (eq. 1), this velocity can be estimated by the following: 
\begin{equation}
\int_{V_{\rm sh}}^{\infty} 4 \pi (v t)^2 \rho_{\rm ej} (v) t dv = M_{\rm CS} \ . 
\end{equation}
Obviously, time ($t$) is cancelled out by the effect of the density decrease ($\rho_{\rm ej} \propto t^{-3}$). This expression reduces to the following: 
\begin{equation}
\exp(-x) (x^2 + 2 x + 2) = \frac{2 M_{\rm CS}}{M_{\rm ej}} \ , 
\end{equation}
where $x = V_{\rm sh} / v_{\rm e}$. The kinetic energy initially contained above $V_{\rm sh}$ will be distributed to the shell, and this is estimated as follows: 
\begin{equation}
E_{0} = \int_{V_{\rm sh}}^{\infty} 2 \pi \rho_{\rm ej} (v) v^{4} t^3 dv \ .
\end{equation} 
In the situation under consideration, the interaction must be adiabatic and the radiation energy loss is negligible. Therefore, $E_0$ is divided into the kinematic energy of the shell ($E_{{\rm K}, 0}$) and the internal energy within it ($E_{{\rm th}, 0}$). The relative fraction for the different channels is dependent on the detailed density structures of the ejecta and the CSM, but in general these two contributions are expected to be comparable as is required by  the conservation of momentum and energy. Hereafter we assume that $E_{{\rm K}, 0} = E_{{\rm th}, 0} = 0.5 E_{0}$ as our fiducial case. 

The initial radiation energy density, therefore the initial temperature ($T_0$), is set by the internal energy content by the following equation (note that $R_{\rm sh} = R_0$ as the initial condition): 
\begin{equation}
4 \pi R_0^3 \frac{\Delta R_{\rm sh}}{R_{\rm sh}} a_{\rm r} T_0^4 = E_{{\rm th}, 0} \ , 
\end{equation}
where $a_{\rm r}$ is the radiation constant. Here we assume that the temperature is uniform within the shell. While this might be a poor approximation in the outermost layer where the expansion to the (nearly) vacuum will introduce a temperature slope in the small mass layer, the bulk of the shocked CSM should be well represented by this characteristic temperature. 

Under the one-zone (or thin shell) approximation, the thermal evolution of the shell, without any additional heating source, is described by the first law of thermodynamics \citep{arnett1980,arnett1982,arnett1996}: 
\begin{equation}
\frac{dE}{dt} + P \frac{dV}{dt} = - L \ , 
\end{equation}
where $E$ is the total internal energy, $P$ is the pressure, $V$ is the volume, and $L$ is the luminosity of the diffused radiation. The equation leads to the following solution: 
\begin{equation}
\left(\frac{T^4 R^4}{T_0^4 R_{0}^4}\right) = \exp \left(-\frac{t_{\rm h} t + t^2/2}{t_{\rm h} t_{\rm d} (0)}\right) \ ,
\end{equation}
where $t_{\rm h} = R_0/V_{\rm sh}$, and $t_{\rm d} (0)$ is the diffusion time scale for the initial condition when the CSM is just swept up ($t = 0$, assuming $t \gg t_{\rm h}$). The diffusion time scale is expressed by $t_{\rm d} \sim \rho \kappa \Delta R_{\rm sh}^2/3 c$, where $\rho$ and $\kappa$ are the density and opacity, respectively. For the thin shell, $\rho = M_{\rm CS}/(4 \pi R_{\rm sh}^2 \Delta R_{\rm sh})$, thus 
\begin{equation}
t_{\rm d} \sim \frac{\kappa M_{\rm CS} \Delta R_{\rm sh}}{12 \pi R_{\rm sh}^2 c} \ .
\end{equation}
Since $L \sim E / t_{\rm d}$ (where $E$ and $t_{\rm d}$ are described as a function of time), the luminosity is described as follow: 
\begin{eqnarray}
L & = & L_0  \exp \left(-\frac{t_{\rm h} t + t^2/2}{t_{\rm h} t_{\rm d} (0)}\right) \ , \\ 
L_0 & = & \frac{E_{{\rm th}, 0}}{t_{\rm d} (0)}\ .
\end{eqnarray}

For the radiation output from the shell, we assume that the spectral energy distribution (SED) is described by the Plank function, which is a reasonable assumption in the situation under consideration. Also, we assume that half of the luminosity is emitted toward the observer, i.e., $L_{\rm obs} = 0.5 L$, to take into account the radiation absorbed by the main ejecta. The radiation temperature ($T_{\rm r}$) can be therefore estimated by the blackbody relation, $L_{\rm obs} = 4 \pi R_{\rm sh}^2 \sigma_{\rm sb} T_{\rm r}^4$, where $\sigma_{\rm sb}$ is the Stefan-Boltzmann constant, as the photosphere position is always close to $R_{\rm sh}$ in the early phase. The multiband light curves can then be calculated by convolving a filter function of each band pass to this blackbody (BB) SED. 

Note that the bolometric luminosity decreases monotonically in time. Peaks are created in different band passes due to the increasing peak wavelength of the SED (which is a result of the `cooling', i.e., decreasing photon temperature). This is the situation analogous to the post shock breakout cooling emission from CCSNe \citep[e.g.,][]{arnett1980,bersten2012}. 

\subsection{Interaction with a Companion Star}

The situation for the hydrodynamical interaction with a companion star can be described in a manner similar to the case for the CSM interaction. The main difference is that the situation here is highly asymmetric by nature, and the interaction creates a hole within a certain solid angle down to the deepest part of the ejecta \citep{marietta2000}, rather than a spherically distributed (high velocity) thin shell. \citet{kasen2010} developed approximate analytical formulae to describe the general properties of the expected radiation, which were used to explain the radiation transfer simulation results. 

In this section, we construct a simple model to describe the observational outcome. Our approach is similar to the one developed for the CSM interaction (\S 2.1). While we share the physical processes same with those adopted by \citet{kasen2010}, details differ; our approach is based on global properties of the interaction, while \citet{kasen2010} based his analysis on local properties of the interaction. As we will show in \S 3.2, the two approaches indeed provide reasonably similar results within the accuracy of interest in this paper. We note that our derivation relies on the hypothesis that the time scale for the equipartition between the thermal (gas) energy created by the interaction and the photon energy is sufficiently short, which is questioned by \citet{kutsuna2015} (see below). 

Unlike the CSM interaction, the mass of the `target' (i.e., the companion star) generally exceeds the mass of the ejecta within the corresponding solid angle. For example, for the solid angle of $\Omega/4 \pi \sim 0.1$, the ejecta mass in this region is $\sim 0.14 M_\odot$ (for $M_{\rm ej} = 1.4 M_\odot$) while the companion star has the mass of $\gsim 1 M_{\odot}$. This means that nearly all the ejecta materials enclosed in this solid angle are shocked by the interaction. Therefore, the available energy budget is $E_0 \sim (\Omega / 4 \pi) E_{\rm ej}$. The fraction of this energy going to the thermal energy will also be smaller than the case of the (spherical) CSM interaction, since the shocked material can expand laterally to form an oblique shock, and a part of the material can expand back to the hole created by the interaction to partially refill the hole. This results in a larger fraction of the kinetic energy converted back from the dissipated SN kinetic energy budget than the case of the spherical CSM interaction. We assume $E_{{\rm th}, 0} = 0.25 E_{0}$ for the shocked ejecta material, similar to the estimate by \citet{kasen2010} (their equation 11). 

The configuration we assume is therefore the thermal energy at the interaction ($E_0$) distributed in the ejecta within the solid angle of $\Omega/4 \pi$. For simplicity, we treat this region as a homogeneous cone, similar to the case of the CSM interaction (a homogeneous shell). For the exponential ejecta structure (eq. 1), originally half of the kinetic energy is contained below $V_{0} \sim 4.5 v_{\rm e}$, and therefore we assume that the outer boundary of this homogeneous fireball to be at $V_0$ as a rough estimate. For $M_{\rm ej} = 1.4 M_{\odot}$ and $E_{\rm ej} = 1.5 \times 10^{51}$ erg, $V_0 \sim 13,000$ km s$^{-1}$. This is consistent with the result of the hydrodynamic simulations \citep[e.g.,][]{kutsuna2015} in the distribution of the shocked companion envelope and the shocked ejecta \citep[see figure 1 of][]{maeda2014}. The initial radiation temperature is then given as follows: 
\begin{equation}
\frac{\Omega}{4 \pi} \frac{4 \pi D^3}{3} a_{\rm r} T_0^4 = E_{{\rm th}, 0} \propto \frac{\Omega}{4 \pi} E_{\rm ej} \ , 
\end{equation}
where $D$ is the separation of the binary. Note that $T_0$ is not dependent on the solid angle, as both the energy budget and the volume of the cone have the same dependence to the solid angle (i.e., equivalent to a corresponding isotropic spherical model). 

The evolution of the temperature and luminosity follows the same equations as the case of the CSM interaction (\S 2.1), once the definition for some physical quantities are modified to take into account the differences between the two situations. Therefore, similarly to the CSM interaction scenario, the bolometric luminosity decreases monotonically with time. As for the diffusion time scale, we need the information about the density in the hole. \citet{kasen2010} introduced a simplified ejecta structure after the interaction, in which the hole has the density by a factor of $\sim 10$ smaller than the freely expanding ejecta. This region is covered by a layer of high density and then by the original (unshocked) ejecta structure in the other directions. We adopt the similar approximation and assume that the density is by a factor of $10$ smaller in the hole than the original ejecta without interaction. Therefore, we adopt the diffusion time scale as follows: 
\begin{equation}
t_{\rm dif} = \frac{f_{\rm h} \kappa M_{\rm ej}}{\beta c R_0} \ ,
\end{equation}
where $f_{\rm h} = 0.1$. Here, $\beta = 13.8$ provides a good approximation for a homogeneous (conical) sphere or similar configuration \citep{arnett1996}. Note that this equation is consistent with the one used for the CSM interaction case, if the geometrical factors are appropriately transformed between the two. The estimate of the {\em isotropic} bolometric luminosity is then given by the same expressions as in equations 9 and 10, once $E_{{\rm th}, 0}$ is replaced by the isotropic value (i.e., $4 \pi/\Omega \times E_{{\rm th}, 0}$). The bolometric luminosity ($L$) is then converted to the observed luminosity ($L_{\rm obs}$) by taking into account the fraction of the radiation escaping toward the hole direction within the solid angle (with the remaining fraction absorbed by the main ejecta), which is estimated by the relative fraction of the base area of the cone (toward an observer) to the total area including the side area (toward the main ejecta). We note that the synthesized light curves are for an observer in the direction of the hole (i.e., the companion direction); this will be fainter or even blacked out in the other directions.

Finally, the characteristic (blackbody) temperature ($T_{\rm r}$) can be estimated once the photospheric radius is known. To the first approximation, we could simply adopt $L_{\rm obs} = 4 \pi V_0^2 t^2 \sigma_{\rm sb} T_{\rm r}^4$. A better estimate can be obtained by taking into account the recession of the photosphere. By adopting the original (exponential) ejecta structure but with the reduction of the density by a factor of $f_{\rm h}$, we can estimate the position of the photosphere as a function of time by the following condition: 
\begin{equation}
\int_{v_{\rm ph}}^{\infty} f_{\rm h} \kappa \rho_{\rm ej} (v)  t dv \sim 1 \ . 
\end{equation} 
This expression reduces to the following: 
\begin{equation}
v_{\rm ph} = v_{\rm e} \log\left(\frac{f_{\rm h} \kappa M_{\rm ej}}{8 \pi v_{\rm e}^2 t^2}\right) \ .
\end{equation}
We then estimate $T_{\rm r}$ by $L_{\rm obs} = 4 \pi v_{\rm ph}^2 t^2 \sigma_{\rm sb} T_{\rm r}^4$. By convolving the expected BB SED with the filter band passes, we obtain multi-color light curves. 

The above formalism assumes that the equipartition between the gas and radiation is quickly reached, so that the shock dissipated energy is efficiently used for radiation output. This assumption may not necessarily be realized, and then the expected luminosity by the ejecta-companion interaction will be lower \citep{kutsuna2015}. Indeed, there could be an additional source of radiation which may set the lower limit for the expected early-phase luminosity. With the hole created by the interaction, a fraction of the inner $^{56}$Ni-rich region would be exposed to an observer toward the same direction. The characteristic time scale, set by the diffusion through the rarefied hole, is expressed as follows: 
\begin{eqnarray}
t_{\rm max} & \sim & \sqrt{t_{\rm h} t_{\rm dif}} \nonumber\\
& \sim & 2.6 \ {\rm days} \left(\frac{f_{\rm h}}{0.1}\right)^{0.5} \left(\frac{\kappa}{0.1 {\rm cm}^2}\right)^{0.5} \nonumber\\
& & \left(\frac{M_{\rm ej}}{1.4 M_\odot}\right)^{0.5} \left(\frac{v_0}{13,000 {\rm km s}^{-1}}\right)^{-0.5} \ .
\end{eqnarray}
At this phase the dominant power source is the decay of $^{56}$Ni to $^{56}$Co, for which the (isotropic) luminosity can be roughly estimated as 
\begin{equation}
L(^{56}{\rm Ni}) \sim 3.4 \times 10^{42} {\rm erg \ s}^{-1} \left(\frac{0.1}{f_{\rm h}}\right) \left(\frac{M^{56}{\rm Ni})}{0.6 M_\odot}\right) \ ,
\end{equation}
where $M$($^{56}$Ni) is the total mass of $^{56}$Ni in the SN ejecta. Here, we estimate that the fraction of $^{56}$Ni which participates in the early emission is $f_{\rm h}$, i.e., the same factor for the density within the hole as compared to that in the unshocked ejecta. This is consistent with the picture we use throughout this section for the computation of the companion interaction, since the remaining fraction of the SN material which was originally in the hole direction is assumed to form the high density region covering the hole. The expected bolometric luminosity from this $^{56}$Ni contribution is $\sim -17.5$ mag in the first few days. Using equation 14 and the BB relation, we estimate that the characteristic radiation temperature is $\sim 20,000-25,000$ K at $\sim 2-3$ days after the explosion. 

In the following sections, we mainly discuss the radiation from the dissipated/thermalized kinetic energy assuming the full thermalization, as this is the frequently adopted model to compare to observations. The contribution from the inner $^{56}$Ni will also be dependent on the highly uncertain $^{56}$Ni distribution after the interaction, investigation of which is beyond the scope of the present work. Whenever necessary we will briefly discuss the contribution from $^{56}$Ni in the rarefied hole region. 

\subsection{He Detonation Scenario}

\subsubsection{Main Characteristics}

The He detonation scenario has been investigated mainly in the context of the double detonation model \citep{fink2010,woosley2011}, while recently an increasing attention has be drawn for the He-ignited violent merger model \citep{guillochon2010,pakmor2013,shen2017}. In typical double detonation models, the mass of the He shell spans from a few $10^{-3} M_\odot$ for the Chandrasekhar-mass WD to a few $0.1 M_\odot$ for the sub-Chandrasekhar-mass WD (down to $\sim 0.7 M_{\odot}$ in the WD mass). The mass fractions of different He detonation products are first dependent on the WD mass. For the `minimal' He shell mass to ignite the He detonation in the model sequence of \citet{fink2010} \citep[see also][]{kromer2010}, the resulting mass fraction of $^{56}$Ni spans from 0.01 (for the WD mass of $\sim 0.8 M_{\odot}$) to 0.2 (for $1.4 M_\odot$). The mass fractions of $^{52}$Fe and $^{48}$Cr show smaller variation within $0.05 - 0.1$. There could be additional uncertainties both in the WD evolution to set up the property the He shell and in the treatment of the He detonation numerically. For example, \citet{woosley2011} reported much larger $^{56}$Ni mass fraction and smaller $^{48}$Cr mass fraction than \citet{fink2010}. Table 1 illustrates typical mass fractions found in these works, which are to be used as an input to our radiation transfer simulations in this paper (see \S 2.3.2 for more details on the model description). The models adopting the mass fractions representing the nucleosynthesis products of \citet{fink2010} are denoted as the He detonation sequence A, while those adopting the results of \citet{woosley2011} are denoted as the He detonation sequence B. 

There is clear difference in the physical processes involved in the He detonation scenario than the interaction with CSM or a companion star. For the diffusion of the thermal energy powered by radioactive decays (i.e., the He detonation products), a bolometric light curve initially rises to the peak, then decreases by a combined effect of the photon diffusion and the decreasing radioactive input. This is in contrast to the behavior expected from the interaction case, which predicts that the bolometric luminosity should always decrease while a peak in each band is attributed to the decreasing photon temperature. The typical time scale to reach to the maximum luminosity in the He detonation scenario can be estimated by equating the expansion time scale and the diffusion time scale. This is analogous to the usual $^{56}$Ni/Co-powered model for the main light curves of SNe Ia \citep{arnett1982,arnett1996}. By adopting the expression for the diffusion time scale similar to equation 12 (as the shell can be geometrically thick, but with $f_{\rm h} = 1$), we obtain the following approximated expression: 
\begin{eqnarray}
t_{\rm max} & \sim & 0.8 \ {\rm days} \left(\frac{\kappa}{0.1 \ {\rm cm}^2}\right)^{0.5} \left(\frac{M_{\rm He}}{0.02 M_\odot}\right)^{0.5} \nonumber\\
& & \left(\frac{V_{\rm He}}{20,000 \ {\rm km s}^{-1}}\right)^{-0.5} \ .
\end{eqnarray} 
Note that this estimate is by a factor of $\sim 2-3$ smaller than that we estimated in \citet{jiang2017}, reflecting the uncertainty in analytically estimating the diffusion time scale. 

The dominant power source can be either of the decay chains of $^{56}$Ni/Co/Fe or $^{52}$Fe/Mn/Cr (with an additional input from $^{48}$Cr/V/Ti), depending on the strength of the He detonation, which is further dependent on the WD mass and the mass of the He shell \citep{fink2010,woosley2011}. Assuming that $^{56}$Ni is a main power source, the radioactive energy input is given by 
\begin{eqnarray}
L (^{56} {\rm Ni}) & \sim & 7.8 \times 10^{41} {\rm erg \ s}^{-1} \left(\frac{X_{56}}{0.5}\right) \left(\frac{M_{\rm He}}{0.02 M_{\odot}}\right) \nonumber\\
& & \exp \left(-\frac{t}{8.8 {\rm days}}\right) \ . 
\end{eqnarray}
The radioactive decay chains of $^{52}$Fe/Mn/Cr and $^{48}$Cr/V/Ti and their energy generation rates are summarized in Appendix. The radioactive inputs by these decays are given as follows: 
\begin{eqnarray}
L (^{52} {\rm Fe}) & \sim & 7.6 \times 10^{42} {\rm erg \ s}^{-1} \left(\frac{X_{52}}{0.1}\right) \left(\frac{M_{\rm He}}{0.02 M_{\odot}}\right) \nonumber\\
& & \exp \left(-\frac{t}{0.5 {\rm days}}\right) \ . 
\end{eqnarray}
\begin{eqnarray}
L (^{48} {\rm Cr}) & \sim & 3.0 \times 10^{41} {\rm erg \ s}^{-1} \left(\frac{X_{48}}{0.1}\right) \left(\frac{M_{\rm He}}{0.02 M_{\odot}}\right) \nonumber\\
& & \exp \left(-\frac{t}{1.3 {\rm days}}\right) \ . 
\end{eqnarray}
At $t \gg 1.3$ days, the dominant power from this chain is replaced by the decay of the daughter nucleus $^{48}$V: 
\begin{eqnarray}
L (^{48} {\rm V}) & \sim & 1.3 \times 10^{41} {\rm erg \ s}^{-1} \left(\frac{X_{48}}{0.1}\right) \left(\frac{M_{\rm He}}{0.02 M_{\odot}}\right) \nonumber\\
& & \exp \left(-\frac{t}{23.0 {\rm days}}\right) \ . 
\end{eqnarray}
At $t \sim 1$ day, the energy input by these radioactive nuclei would be $\sim 7 \times 10^{41}$ erg s$^{-1}$ for $^{56}$Ni (for the mass fraction of 0.5) or $\sim 1 \times 10^{42}$ erg s$^{-1}$ for $^{52}$Fe (for the mass fraction of 0.1). In either case, the expected luminosity is similar. 

Note that the above equations provide the total energy input in the form of decay $\gamma$-rays and positrons (see Appendix). It is equal to the input to the  (UV/optical) bolometric luminosity only if full thermalization is realized. For $V_{\rm He} = 20,000$ km s$^{-1}$ (the He shell velocity) and $M_{\rm He} = 0.02 M_\odot$ (the He shell mass), the optical depth to the $\gamma$-rays within the He layer, adopting the well-defined opacity of $\kappa_{\gamma} = 0.027$ cm$^{-2}$ g$^{-1}$ as determined by the Compton scatterings, is $\sim 3$ at one day after the explosion. Therefore, full thermalization within the He layer is realized around the `early peak' in most of the model sequence, and the above equations provide good estimate of the (UV/optical) bolometric luminosity. The approximation is worse for the models with a thinner He layer, and a non-negligible amount of the $\gamma$-ray leakage should be taken into account especially for the most massive WD models in our model sequence (with the mass of the He layer being $3.5 \times 10^{-3} M_\odot$). We note that this effect is indeed taken into account in our radiation transfer simulations (\S 3 and thereafter). 

We expect that the photosphere is within the He-rich surface region in the first few days, covering the (first) peak. Therefore, the first-order approximation is $L_{\rm bol} = 4 \pi V_{\rm He}^2 t^2 \sigma_{\rm sb} T_{\rm r}^4$. This provides the estimate of the radiation temperature as $T_{\rm r} \sim 7,000 - 15,000$K at $t \sim t_{\rm max} \sim 1-4$ day, where $V_{\rm He}$ is the characteristic velocity of the He layer ($\sim 20,000$ km s$^{-1}$), for the bolometric luminosity of $\sim 10^{42}$ erg s$^{-1}$. This suggests that the radiation created by the He detonation can peak in the optical wavelength. 

Some of the expected characteristic behaviors would be shared by the `inner $^{56}$Ni contribution' in case of the companion interaction (\S 2.2). The main difference would be the radiation temperature and SED. For the `innerf $^{56}$Ni contribution within the hole, due to the confinement of $^{56}$Ni in the innermost region, the effect of the recession of the photosphere is expected to be significant, leading to the high temperature ($\gsim 20,000$K; see \S 2.2). 

\subsubsection{Models and Method for Radiation Transfer Simulations}

Given the complications involved in the He detonation scenario, the details will be discussed based on radiation transfer simulations, rather than analytic expressions. For the input models to the radiation transfer simulations, we construct a series of toy (kinetic) models, which take into account main features found in the hydrodynamic simulations of the double detonation models \citep{fink2010,kromer2010}. The procedures to construct the model are the same as in \citet{jiang2017}, and we refer Extended Figures 5 and 6 of \citet{jiang2017} which illustrate the kinematic and composition structures adopted in our models.

Our model sequence is summarized in Tables 1 and 2. Our main parameters describing the configuration of the explosion outcome are the following: the mass of the progenitor WD, the chemical composition (and the mass) of the burnt He layer. The model is named, for example, 12A, where the first digits describe the WD mass ($1.2 M_\odot$ for this example), and the character (A or B) denotes the composition type in the He layer (see \S 2.3.1 and Table 1). Additional models are denoted by `N' which have the same main SN ejecta structure but without the He shell. In addition, we examine two different configurations for a Chandrasekhar-mass WD progenitor; pure-detonation and W7-like model. The carbon detonation in a Chandrasekhar-mass WD predicts $M$($^{56}$Ni) $\gsim 1 M_{\odot}$. Additionally, the W7-like model is constructed by (artificially) setting $M$ ($^{56}$Ni) to be $0.61 M_{\odot}$ (a typical amount of $^{56}$Ni in normal SNe Ia). The latter case is denoted by additional character, `L' (from `Lower luminosity'), in the end of the model name. This may happen either in the He accreting scenario in the SD or the He-ignited violent merger scenario in the DD. In the former, the shock wave penetrating deep in the C+O WD may not become sufficiently strong to initiate the central detonation for a near Chandrasekhar-mass WD. In the latter, if a significant amount of the companion WD would also be disrupted by the merger and swept up by the SN ejecta\footnote{In practice, this is unlikely since the He-ignition will take place before the companion WD is tidally disrupted \citep{pakmor2013,tanikawa2015}}, the primary WD mass could be smaller than the Chandrasekhar mass (note that the ejecta mass in our model is a sum of the primary WD and additional mass from the companion WD, if interpreted as the He ignited violent merger model). 

\begin{deluxetable}{cccc}
 \tabletypesize{\scriptsize}
 \tablecaption{He Detonation Sequence\tablenotemark{a} ($M_\odot$)
 \label{tab2}}
 \tablewidth{0pt}
 \tablehead{
   \colhead{Models}
 & \colhead{$M$($^{56}$Ni)}
 & \colhead{$M$($^{52}$Fe)}
 & \colhead{$M$($^{48}$Cr)}}
\startdata
8A & 8.4e-4 & 7.6e-3 & 1.1e-2\\
9A & 1.1e-3 & 7.0e-3 & 7.8e-3\\
10A &1.7e-3 & 6.2e-3 & 4.4e-3\\
11A & 4.4e-3 & 3.5e-3 & 2.2e-3\\
13A & 1.5e-3 & 1.2e-3 & 6.8e-4\\
14A/AL & 5.7e-4 & 2.0e-4 & 1.5e-4\\
8B & 6.3e-2 & 7.6e-3 & 2.2e-3\\
9B & 4.2e-2 & 7.0e-3 & 1.6e-3\\
10B &2.8e-2 & 6.2e-3 & 8.8e-4\\
11B & 2.0e-2 & 3.5e-3 & 4.4e-4\\
13B & 6.5e-3 & 1.2e-3 & 1.4e-4\\
14B/BL & 1.8e-3 & 2.0e-4 & 3.0e-5\\
\enddata
\tablenotetext{a}{Other elements (Ti, Ca, Ar, S, Si, Mg, O, C, He) are also included. The composition is dominated by He in the sequence A. In the sequence B, $^{56}$Ni is the dominant species (with the mass fraction fixed to be 0.5), followed by He.}
\end{deluxetable}

\begin{deluxetable}{cccc}
 \tabletypesize{\scriptsize}
 \tablecaption{SN Ejecta Models ($M_\odot$)
 \label{tab1}}
 \tablewidth{0pt}
 \tablehead{
   \colhead{Models}
 & \colhead{WD mass}
 & \colhead{He shell mass\tablenotemark{a}}
 & \colhead{$M$($^{56}$Ni)}}
\startdata
8A/B/N & 0.81 & 0.126 & 0.17 (A/N) or 0.37 (B)\\
9A/B/N & 0.92 & 0.084 & 0.34 (A/N) or 0.54 (B)\\
10A/B/N & 1.03 & 0.055 & 0.55 (A/N) or 0.75 (B)\\
11A/B/N & 1.13 & 0.039 & 0.78 (A/N) or 0.98 (B)\\
13A/B/N & 1.28 & 0.013 & 1.05 (A/B/N)\\
14A/B/N & 1.38 & 0.0035 & 1.10 (A/B/N)\\
14AL/BL/NL & 1.38 & 0.0035 & 0.61 (AL/BL/NL)\\
\enddata
\tablenotetext{a}{The He shell mass is set to be zero for the sequence N.}
\end{deluxetable}

We adopt the exponential density structure (eq. 1). The ejecta mass ($M_{\rm ej}$) is determined as a sum of the WD mass and the He shell mass. Once the masses of the nucleosynthesis products are specified, it is straightforward to compute the nuclear energy generation. By subtracting the nuclear energy generation by the binding energy of the WD (and that of the He layer), we obtain the final kinetic energy. For the binding energy, we compute it for different WD masses with zero temperature\footnote{The binding energy of the cold WD was computed with a code developed by Frank Timmes 
(http: \slash \slash cococubed.asu.edu \slash code\_pages \slash codes.shtml).}.

For the abundance stratification, we consider (a) the layered structure within the C+O WD and (b) the He detonation in the surface He layer. For (a), we divide the layers to (a1) electron capture-zone, (a2) $^{56}$Ni-zone, (a3) IME-zone, and (a4) O-zone. For (b), we assume a single chemical zone structure representing the typical He detonation composition, following the results by \citet{fink2010}. In the classical double detonation model, we set the mass of the electron capture-zone zero. Note that the zone (a1) is introduce only for the Chandrasekhar WD model (as a reference, for Models 14AL, 14BL, and 14NL), but the mass there is set to be zero in the other models. 

As for the abundances in the core burning and in the He layer, we examine two cases: sequence A (the weaker He detonation) and sequence B (the stronger He detonation). These are constructed by inspecting the outcome of the He detonation simulations by \citet{fink2010} (for sequence A) and that in \citet{woosley2011} (for sequence B). Specifically, the model sequence A is constructed so that the masses of each species ($^{56}$Ni, $^{52}$Fe, $^{48}$Cr, Ti, Ca, Ar, S, Si, Mg, O, C) are the same with the results from \citet{fink2010} \citep[see also][]{kromer2010} both in the main SN ejecta and in the He layer. For the He detonation products, the remaining fraction is assumed to be He, and it indeed dominates the composition in the He layer. For the sequence B, we assume a fixed mass fraction of $^{56}$Ni to be 0.5 in the He layer. The mass fraction of $^{48}$Cr is reduced by a factor of 5 from the sequence A. The change in the total mass is then compensated by changing the He mass fraction to conserve the total mass in the He layer. In the sequence B, the dominant species in the He layer is $^{56}$Ni, followed by He. The mass of $^{56}$Ni in the main SN ejecta is also increased by $0.2 M_\odot$ in the sequence B, unless it exceeds $1 M_\odot$. The main difference between the two sequences can be seen in Table 1. 

In sum, our model is specified by the WD mass, He shell mass, and the pattern of the He detonation products (A or B). For further simplicity, we test in this paper the `minimal He shell' models, which provides the minimal mass of the He shell required to initiate the strong detonation\footnote{But see \citet{shen2014}, who argue that the minimal He shell mass can be further decreased if a significant amount of carbon is mixed into the He shell. Such cases would become to resemble the pure carbon detonation models without the He shell, similar to the model sequence N \citep[see also][]{sim2010,woosley2011,blondin2017,shen2017}.}. Therefore, our models are mainly specified by the WD mass ($0.8-1.4 M_\odot$), with additional variants for the He detonation products (A, B, and N). 

For the initial kinematic models constructed above, we perform (LTE) multi-frequency and time-dependent Monte Carlo radiation transfer simulations. The details of the radiation processes involved and numerical schemes were presented in \citet{maeda2006} and \citet{maeda2014}. In this work, we updated the code to include the radioactive decays of $^{52}$Fe/Mn/Cr and $^{48}$Cr/V/Ti in addition to $^{56}$Ni/Co/Fe. Note that we do not assume full thermalization of the decay $\gamma$-rays in the simulations (while positrons are assumed to be absorbed in situ); the $\gamma$-ray transfer is solved starting with the $\gamma$-ray lines from the decay chains as mentioned above (see also Appendix), taking into account the photoelectric absorptions, Compton scatterings, and pair productions \citep{maeda2006}.

\section{Early-Phase Light Curves}

\subsection{CSM Interaction}

\begin{figure*}
\begin{center}
\hspace{-2cm}
        \begin{minipage}[]{0.45\textwidth}
                \epsscale{1.6}
                \plotone{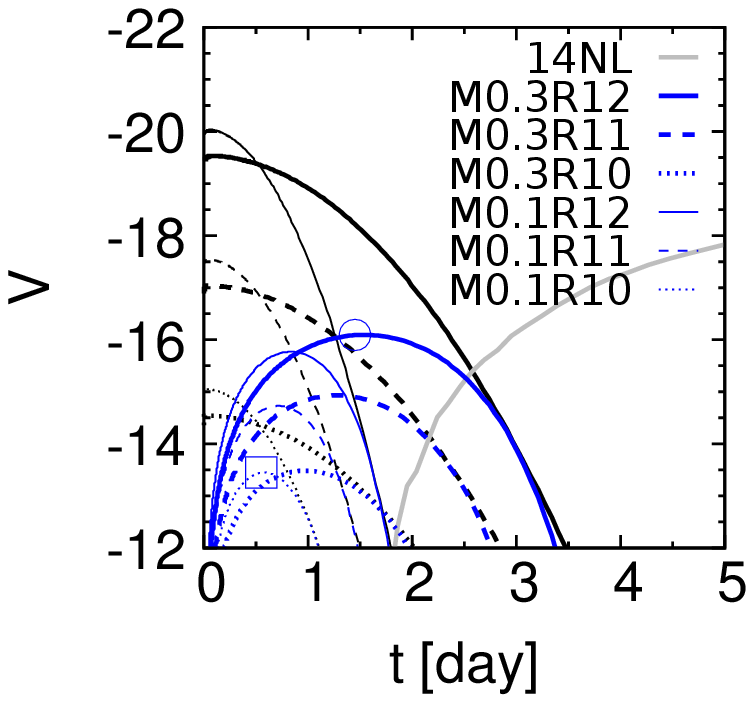}
        \end{minipage}
        \begin{minipage}[]{0.45\textwidth}
                \epsscale{1.6}
                \plotone{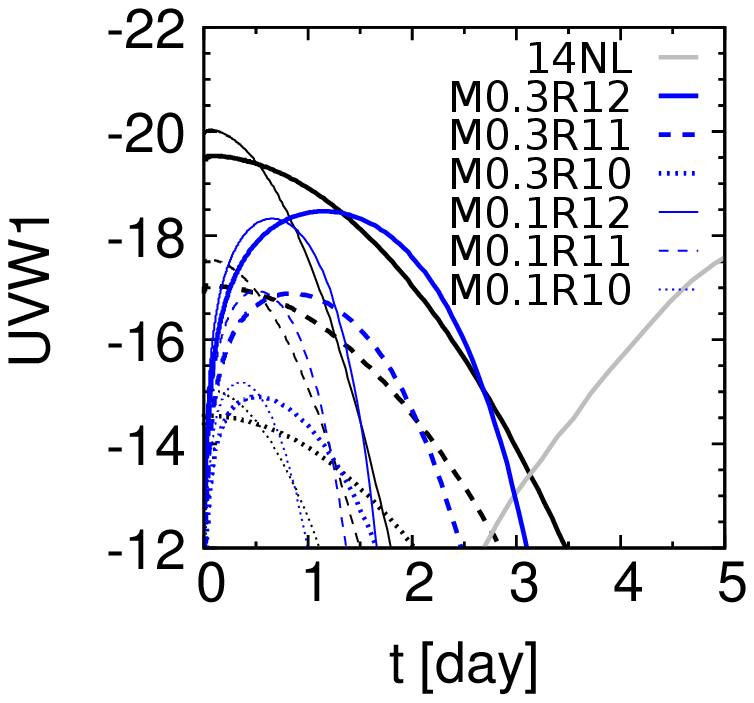}
        \end{minipage}\\
\hspace{-2cm}
        \begin{minipage}[]{0.45\textwidth}
                \epsscale{1.6}
                \plotone{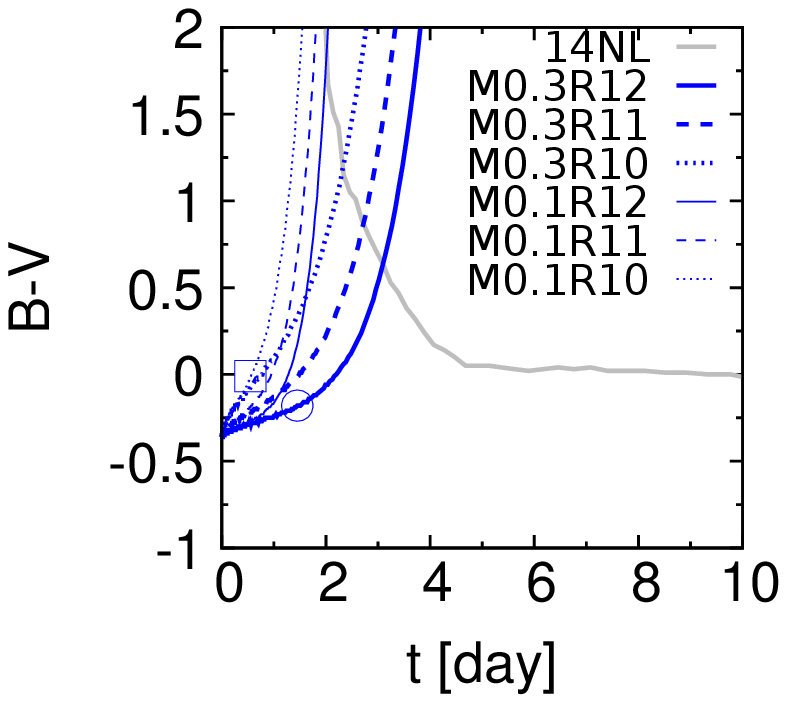}
        \end{minipage}
        \begin{minipage}[]{0.45\textwidth}
                \epsscale{1.6}
                \plotone{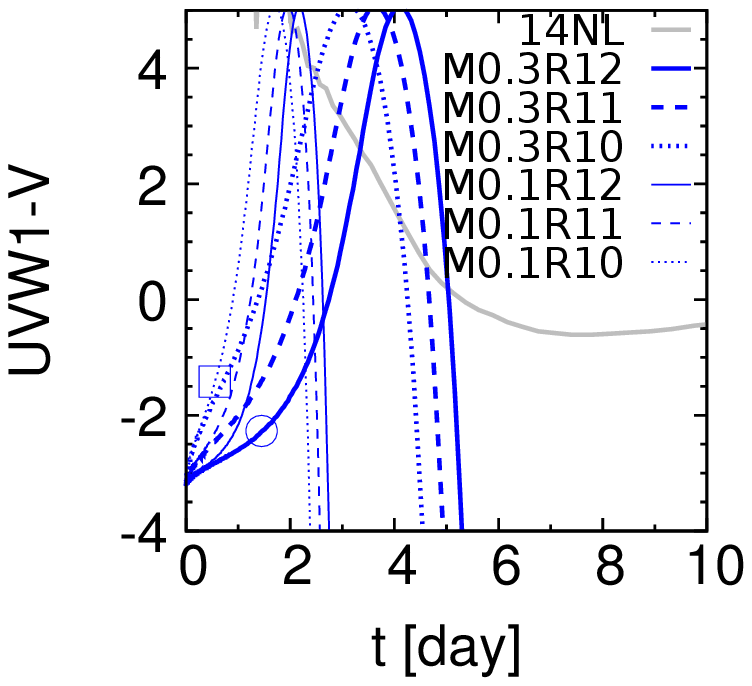}
        \end{minipage}
\end{center}
\caption
{Simulated light curves and color evolutions from the CSM interaction models. The model parameters are the CSM mass ($0.3 M_{\odot}$ shown by thick lines and $0.1 M_{\odot}$ shown by thin lines) and the characteristic (outer edge) radius of the CSM ($10^{12}$, $10^{11}$, $10^{10}$ cm, as shown by solid, dashed, and dotted lines, respectively). The $V$-band (top-left panel) and $UVW1$-band (top-right) light curves are shown by blue lines, while the bolometric light curves are shown by black lines. Also shown are the light curves of the model 14NL in each band (gray), which illustrates a typical SN light curve powered by (inner) $^{56}$Ni/Co/Fe decays. The $B-V$ and $UVW1-V$ color evolutions are shown in the bottom panels. The peak in the $V$-band for two selected models ($0.3 M_{\odot}$ and $10^{12}$ cm, $0.1 M_\odot$ and $10^{10}$ cm) is indicated by the open circle and square, respectively. The $B-V$ and $UVW1-V$ colors at the corresponding epochs are indicated by the same symbols. 
\label{fig1}}
\end{figure*}

\begin{figure*}
\begin{center}
\hspace{-2cm}
        \begin{minipage}[]{0.45\textwidth}
                \epsscale{1.6}
                \plotone{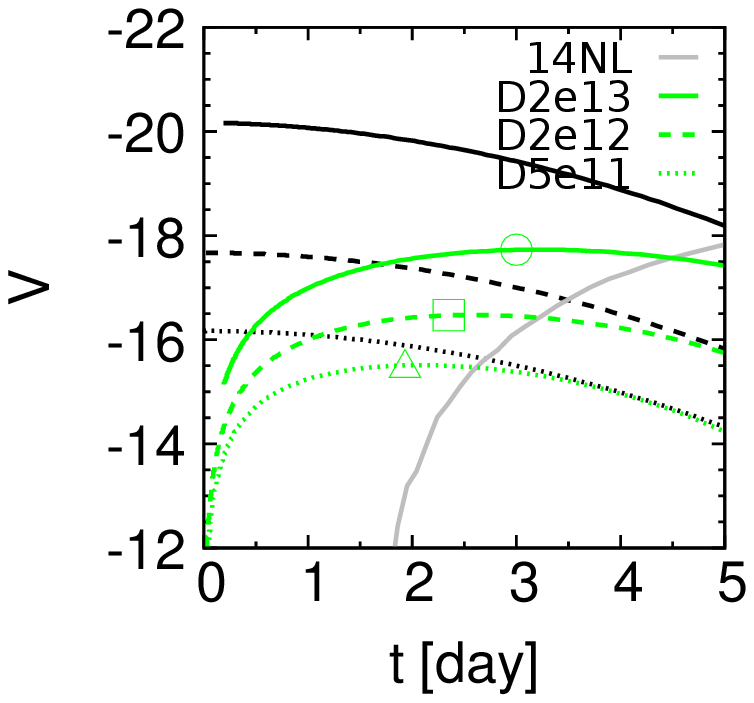}
        \end{minipage}
        \begin{minipage}[]{0.45\textwidth}
                \epsscale{1.6}
                \plotone{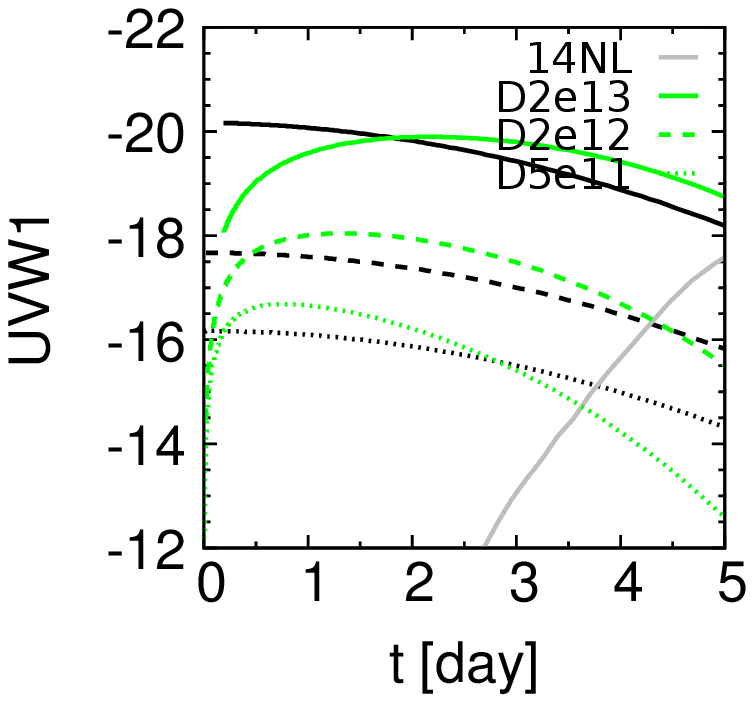}
        \end{minipage}\\
\hspace{-2cm}
        \begin{minipage}[]{0.45\textwidth}
                \epsscale{1.6}
                \plotone{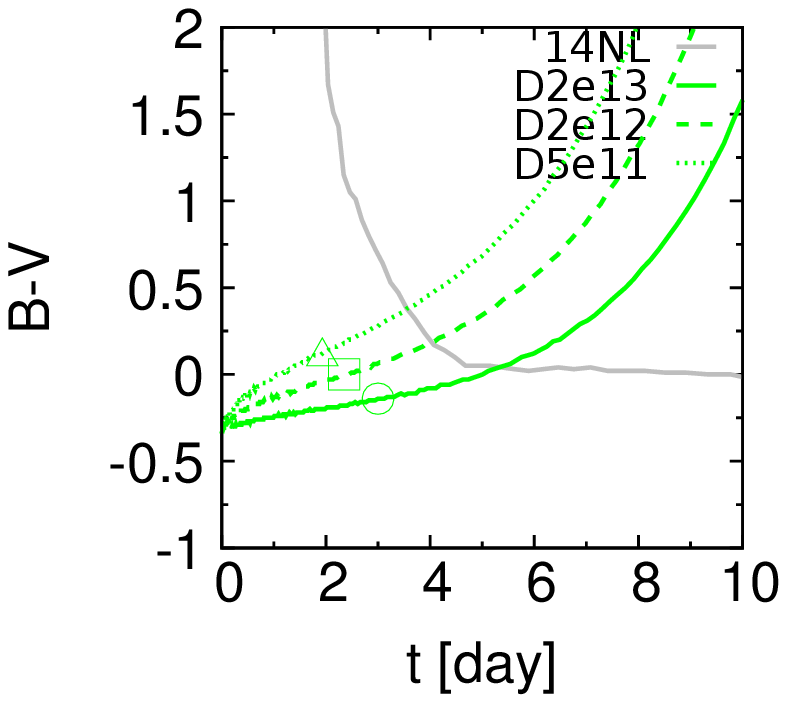}
        \end{minipage}
        \begin{minipage}[]{0.45\textwidth}
                \epsscale{1.6}
                \plotone{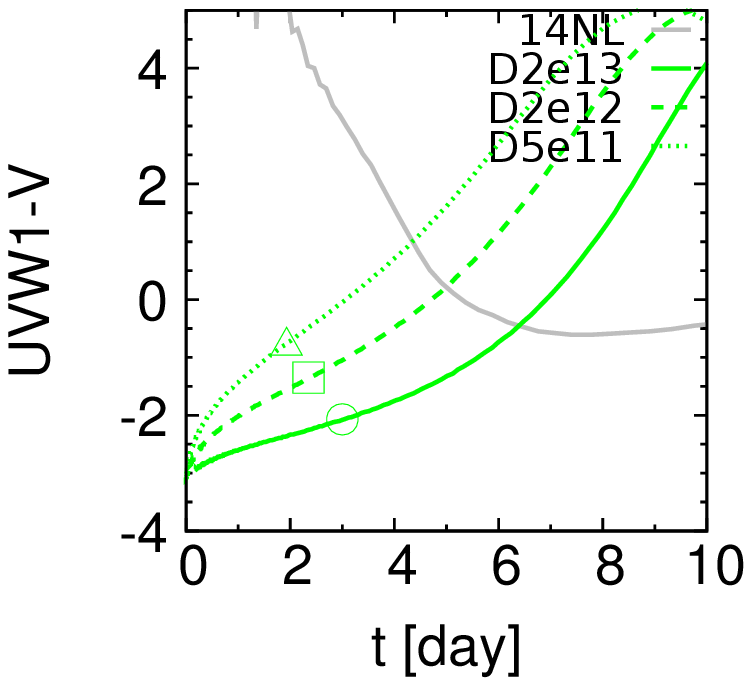}
        \end{minipage}
\end{center}
\caption
{Simulated light curves and color evolutions from the SN-companion interaction models. The model parameter is the separation ($2 \times 10^{13}$, $2 \times 10^{12}$, $5 \times 10^{11}$ cm, as shown by solid, dashed, and dotted lines, respectively). The $V$-band (top-left panel) and $UVW1$-band (top-right) light curves are shown by green lines, while the bolometric light curves are shown by black lines. Also shown are the light curves of the model 14NL in each band (gray), which illustrates a typical SN light curve powered by (inner) $^{56}$Ni/Co/Fe decays. The $B-V$ and $UVW1-V$ color evolutions are shown in the bottom panels. The open circle, square, and triangle show the properties in the $V$-band light curves and the colors at the corresponding $V$-band peak dates. 
\label{fig2}}
\end{figure*}

Figure 1 shows the synthetic light curves (in the $V$-band and the {\em SWIFT} $UVW1$-band)\footnote{In this paper, we use the $UVW1$ band as a representative UV behavior. While the $UVM2$-band provides the cleaner window without the `red leak' in its sensitivity function, the wavelength range is substantially contaminated by the UV line blending effect and involves a large uncertainty in the radiation transfer model such as the line list \citep{pinto2000,kasen2006,kromer2009}.} for the ejecta-CSM interaction, which are computed following the formalisms presented in \S 2.1. The model parameters ($M_{\rm CS}/M_{\odot}, R_0/{\rm cm}$) are the following: ($0.3$, $10^{12}$), ($0.3$, $10^{11}$), ($0.3$, $10^{10}$), and ($0.1$, $10^{12}$), ($0.1$, $10^{11}$), ($0.1$, $10^{10}$), similar to the parameter space examined by \citet{piro2016}. Note that the computation here does not include the main SN ejecta component powered by $^{56}$Ni/Co decay. As a guide, we show synthetic light curves for our fiducial model 14NL (\S 2.3.2, computed through the detail radiation transfer simulations), which is similar to the W7 model. 

Despite our simplified analytic treatment, we find reasonably good agreement to the $V$-band light curves obtained with radiation hydrodynamic calculations by \citet{piro2016}. Our light curve tends to become slightly brighter than the results obtained by \citet{piro2016}. This likely comes from the uncertainties in the fraction of the generated kinetic energy converted to the internal energy or in the opacity (for which we adopt $\kappa = 0.1$ cm$^{-2}$ g$^{-1}$ for the interaction scenarios).  The dependence of the light curve characteristics to the input parameters is similar, suggesting that our model captures the basic physical processes involved in the SN-CSM interaction and resulting radiation output. The early-phase peak in the first few days is predicted to be distinguishable from the SN main light curve powered by $^{56}$Ni, which may however be affected by the behavior of the early rising light curve from the SN main component \citep[depending on the mixing of $^{56}$Ni;][]{piro2012,piro2013,piro2014,piro2016}. Somehow unexpectedly from the naive expectation, we find that in general it can be easier to distinguish the early-phase signature from the CSM interaction for more confined/compact CSM distribution (i.e., smaller $R_0$), despite the fainter peak magnitude. The earlier emergence helps to separate this component from the main SN emission. 

The simulated $UV$ light curves (for the {\em SWIFT} $UVW1$-band) are shown in Figure 1. In the $UV$, the light curves peak slightly earlier than in the optical, and the peak luminosities are much larger than in the optical. Given the deficiency of the $UV$ light from the $^{56}$Ni-powered SN ejecta due to the line blending by Fe-peaks and heavy elements (see the light curve of Model 14NL in the same figure), the $UV$ provides a much cleaner window than the optical to catch the early flash caused by the CSM interaction. A drawback is the time scale even shorter than in the optical. Catching this signature will require either a high cadence survey in the $UV$ bands or a very quick follow-up in the $UV$ wavelength following the discovery by high cadence optical surveys. 

It should be emphasized that this combined behavior in the optical and $UV$ can be readily understood in terms of the basic radiation process involved. As shown in Figure 1, the CSM interaction predicts a monotonically decreasing bolometric light curve. The peaks in different bands are the result of the increase in the characteristic wavelength of the (essentially BB) radiation with time, which first passes through the $UV$ and then in the optical. This process creates the characteristic behavior in which the flux peaks earlier at larger luminosity, for shorter wavelength. This shares the same behavior to the `cooling emission' predicted and observed for CCSNe (especially those with an extended envelope) \citep[e.g.,][]{arnett1980,bersten2012}, since the main physical processes involved are indeed the same.

The characteristic time scale for the evolution in the bolometric light curves is given by $\sqrt{t_{\rm h} t_{\rm d}(0)}$, after which the decrease in the luminosity and temperature is accelerated due to the photon diffusion (eqs. 7 and 9). This time scale is expressed as $\propto \sqrt{\kappa M_{\rm CS} / V_{\rm sh} (\Delta R_{\rm sh}/R_{\rm sh})}$, thus mainly determined only by $M_{\rm CS}$ (neglecting the details including the dependence of $V_{\rm sh}$ on $M_{\rm CS}$). This is given as $\sim 0.7$ days for $M_{\rm CS} = 0.1 M_{\odot}$, and $\sim 1.3$ day for $M_{\rm CS} = 0.3 M_{\odot}$. Therefore, the bolometric luminosity starts the exponential decline from the very beginning, which also creates the rapid decrease in the temperature. As such, the time scale is quite short, and this is the reason why the difference in the timing of the optical peak and in the $UV$ peak is relatively small, since the shift in the wavelength happens quickly due to the rapid cooling. 

Another thing to note is that a large fraction of the ejecta kinetic energy is converted to the thermal energy in the SN-CSM interaction scenario. For the models with $M_{\rm CS}=0.3 M_\odot$ and $0.1 M_\odot$, $E_0/10^{51} {\rm erg} = 0.63$ ($V_{\rm sh} \sim 12,000 {\rm km s}^{-1})$ and $E_0/10^{51} {\rm erg} =0.34$ ($V_{\rm sh} = 16,000 {\rm km s}^{-1})$, respectively. The behavior is consistent with numerical hydrodynamic simulations \citep[e.g.,][]{noebauer2016}. While the effect of the interaction on the maximum-light spectra is beyond the scope of the present work, this would suggest that a large amount of the CSM would change the line profiles for the main SN ejecta emission before or around the maximum light. This would suppress the absorption with the velocity substantially larger than $V_{\rm CS}$ as compared to the non-interaction model. We may further expect that it could on the other hand enhance the absorption around $V_{\rm CS}$, which has been suggested to explain the high-velocity features \citep{gerardy2004,mulligan2017} \citep[but see][for a different scenario]{zhao2016}. Therefore, it may create very flat evolution in the line velocity evolution as a function of time before or even around maximum-light. This is quite different than the standard prediction without the CSM interaction, and may be in conflict with the early phase data available for a number of SNe, which typically show rapid decrease in the line velocities \citep[e.g.,][]{benetti2005}. 

The color evolution is a powerful probe to the underling mechanism, as this is free from the uncertainty in the extinction, and it is especially useful in investigating the {\em UV} behavior. The $B-V$ color evolves from blue ($B-V \sim -0.3$) and quickly becomes redder as time goes by. At the $V$-band peak, the color is still relatively blue ($B-V \lsim 0$), which is a general behavior expected in the `cooling emission' (see also \S 4). As time goes by, the color evolves further to the red ($B-V > 0$), then the radiation resulting from the CSM interaction is overwhelmed by the rising light curve of the $^{56}$Ni-powered main light curve (see Model 14NL). Then the color becomes increasingly blue again to $B-V \sim 0$, as expected from emission from the $^{56}$Ni-powered main SN ejecta \citep[see also][]{piro2016}.

The {\em UV}-optical color evolution shows a pronounced effect of the interaction. Overall behavior is qualitatively similar to the $B-V$ color, but it starts with a much bluer color and evolves faster than in the optical (Fig. 1). The {\em UV}-optical color is very blue at the moment of the optical peak with $UVW1-V \sim -2$ mag (in the Vega system). The large contrast in the color in the early flash and that in the rising phase by the $^{56}$Ni heating in the main ejecta, e.g., $\sim 2$ magnitude contrast in the first few days and $\sim 10$ days after the explosion, can be a very useful diagnosing observable to test the interaction scenario, as this is free from the uncertainty in the extinction. 

\subsection{Companion Interaction}

Figure 2 shows the light curves for the companion interaction scenario, computed following the formalisms given in \S 2.2. Again, the model curves do not include the main SN component, and the light curves simulated for the $^{56}$Ni-powered model 14NL are shown for demonstration. The pre-SN separation ($D$) between the SN progenitor and the companion star is a main input parameter characterizing the resulting radiation output. Figure 2 shows  the models with $D = 2 \times 10^{13}$ cm, $2 \times 10^{12}$ cm, and $5 \times 10^{11}$ cm. These correspond to the following companion stars filling the Roche Lobe: RG ($\sim 1 M_{\odot}$), MS ($\sim 6 M_\odot$), and MS ($\sim 2 M_{\odot}$), respectively, which are similar to the situations studied by \citet{kasen2010} and \citet{kutsuna2015}. 

We observe good agreement with the results by \citet{kasen2010} in the bolometric luminosities. Our simulated bolometric luminosities are initially $\sim -20$, $-18$, and $-16$ mag, similar to those found by the simulations by \citet{kasen2010}. Furthermore, for example, the model with $D = 2 \times 10^{13}$ cm decays in its luminosity by 1 mag in $\sim 4$ days, fully consistent with \citet{kasen2010}. We also see fair agreement in the $V$-band and $UV$-band properties with the previous study\footnote{Note that we use the Vega system while \citet{kasen2010} uses the AB system.}. 

The basic behavior is the same with the CSM interaction model (\S 3.1), and therefore similar diagnostic observables apply including the importance of the {\em UV}-optical color evolution. However, there are several important differences. First, the companion interaction models presented here lead to longer time scale than the SN-CSM interaction models. The characteristic time scale is expressed as $\propto \sqrt{\kappa M_{\rm ej}/ V_0}$, thus it is the same for all the companion interaction models presented here. Interestingly, within our formalism the characteristic time scale (in the evolution of thermal condition and thus the bolometric light curve evolution) does not depend on the model parameters (i.e., $t_{\rm dif} \propto R_0^{-1}$ and $t_{\rm h} \propto R_0$ cancel out). The exact value can be uncertain (beyond the simple model adopted here), while we expect that this characteristic time scale is larger for the companion interaction case than in the SN-CSM interaction case. 

Another important difference is in its initial internal energy content created by the interaction. Since the energy dissipation is limited by the solid angle of the interaction (eq. 11), it is the same value of $\sim 8 \times 10^{49}$ erg irrespective of the separation ($D$) and other parameters, unlike the CSM interaction case where the CSM mass determines the energy generation. This is much smaller than the SN-CSM interaction where a large fraction of the kinetic energy is thermalized. Combined with the above mentioned (relatively) long time scale, the companion interaction scenario predicts an interesting difference in its behavior than the SN-CSM interaction scenario. The bolometric luminosity decays more slowly than the SN-CSM interaction case. The shift in the peak wavelength can therefore happen before the characteristic time scale (i.e., $\sqrt{t_{\rm dif} t_{\rm h}}$) during the slow evolution (i.e., adiabatic cooling phase), and accordingly this change in the characteristic wavelength is rather gradual. It thus leads to a relatively long delay between the UV peak and the optical peak. Accordingly, the color evolution is also slow as compared to the CSM interaction case (see Fig. 2), while it shares qualitatively similar evolution from the blue to the red. 

\begin{figure*}
\begin{center}
\hspace{-2cm}
        \begin{minipage}[]{0.45\textwidth}
                \epsscale{1.6}
                \plotone{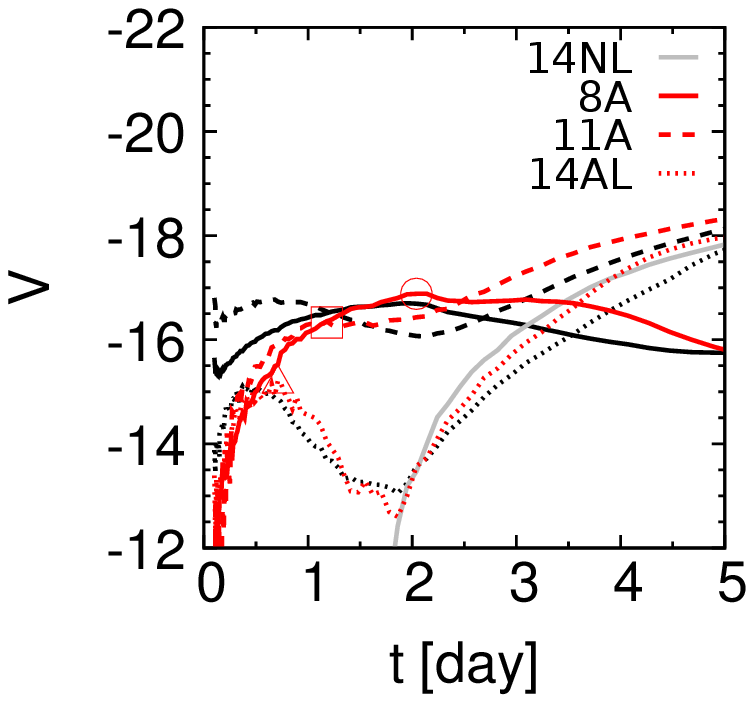}
        \end{minipage}
        \begin{minipage}[]{0.45\textwidth}
                \epsscale{1.6}
                \plotone{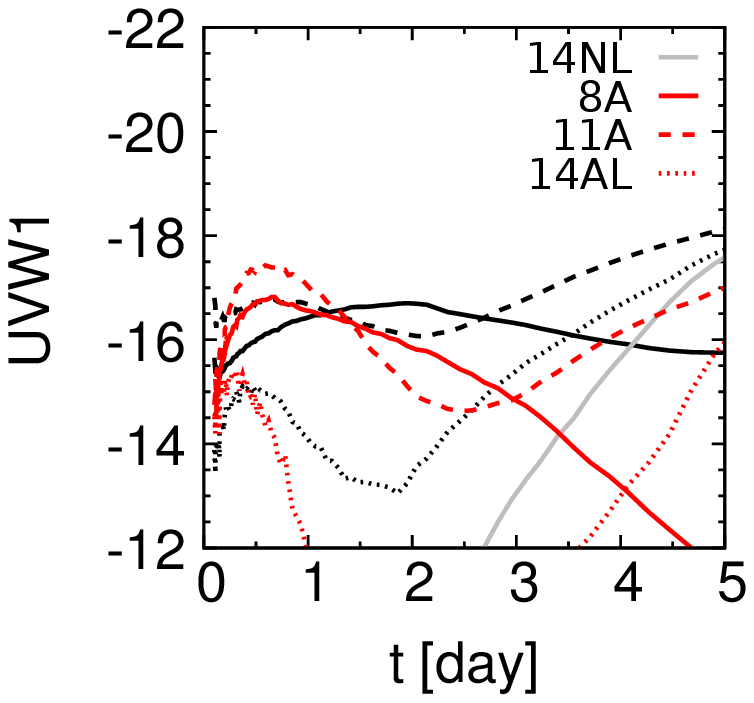}
        \end{minipage}\\
\hspace{-2cm}
        \begin{minipage}[]{0.45\textwidth}
                \epsscale{1.6}
                \plotone{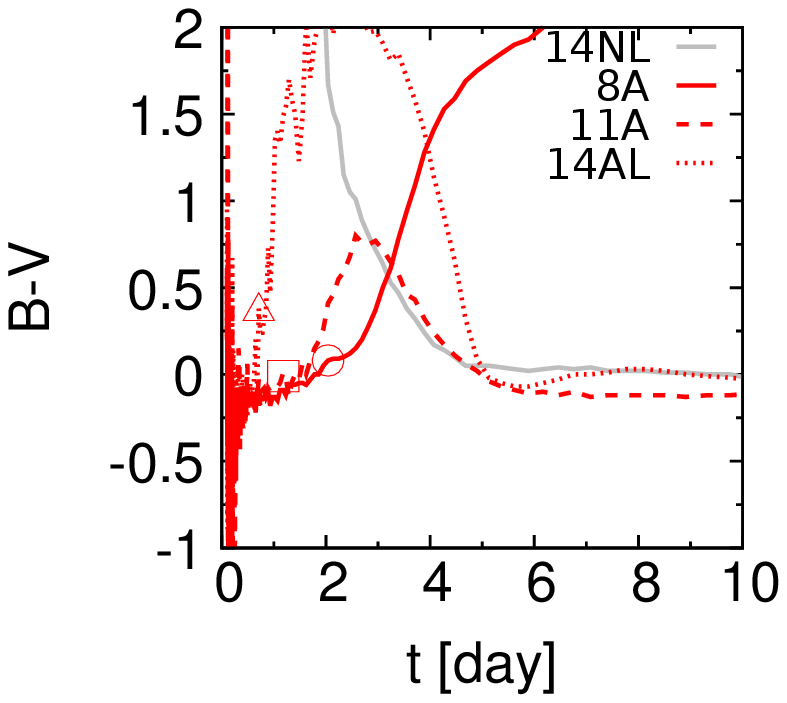}
        \end{minipage}
        \begin{minipage}[]{0.45\textwidth}
                \epsscale{1.6}
                \plotone{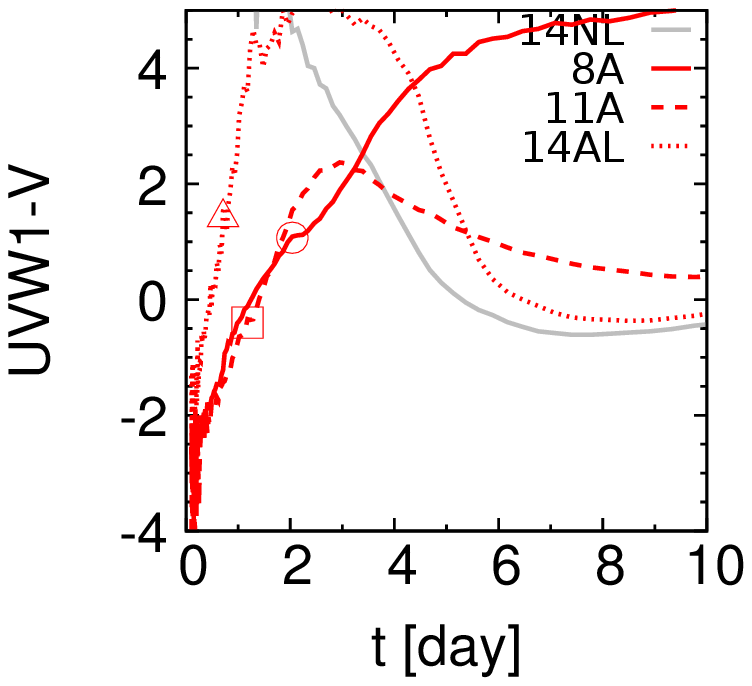}
        \end{minipage}
\end{center}
\caption
{Simulated light curves and color evolutions from the He detonation models (the model sequence A for the nucleosynthesis products). The models shown here are 8A (solid), 11A (dashed), and 14AL (dotted). The $V$-band (top-left panel) and $UVW1$-band (top-right) light curves are shown by red lines, while the bolometric light curves are shown by black lines. Also shown are the light curves of the model 14NL in each band (gray), which illustrates a typical SN light curve powered by (inner) $^{56}$Ni/Co/Fe decays. The $B-V$ and $UVW1-V$ color evolutions are shown in the bottom panels. Note that for the He detonation models, not only the surface layer (the He detonation products) but also the whole SN ejecta (including $^{56}$Ni) is included. The open circle, square, and triangle show the properties in the $V$-band light curves and the colors at the corresponding $V$-band peak dates. 
\label{fig3}}
\end{figure*}

\begin{figure*}
\begin{center}
\hspace{-2cm}
        \begin{minipage}[]{0.45\textwidth}
                \epsscale{1.6}
                \plotone{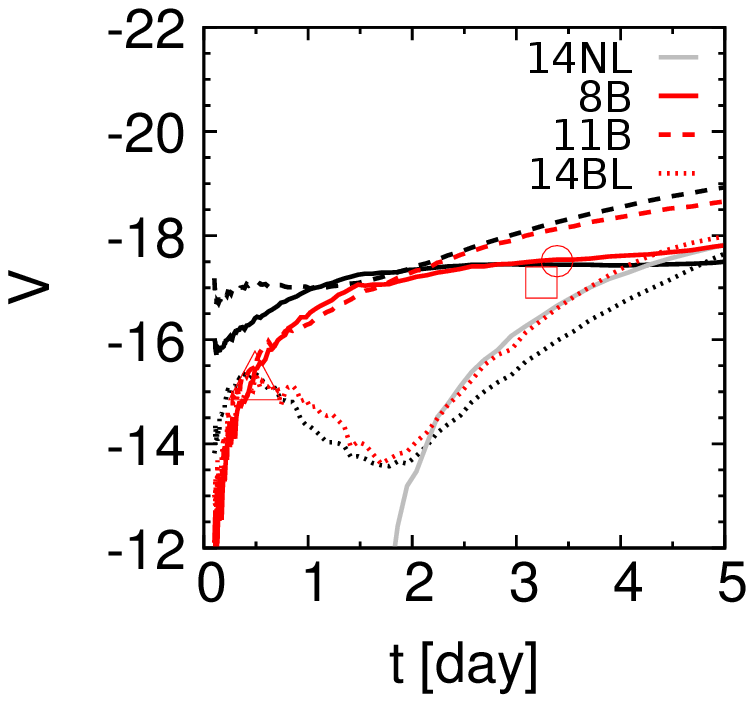}
        \end{minipage}
        \begin{minipage}[]{0.45\textwidth}
                \epsscale{1.6}
                \plotone{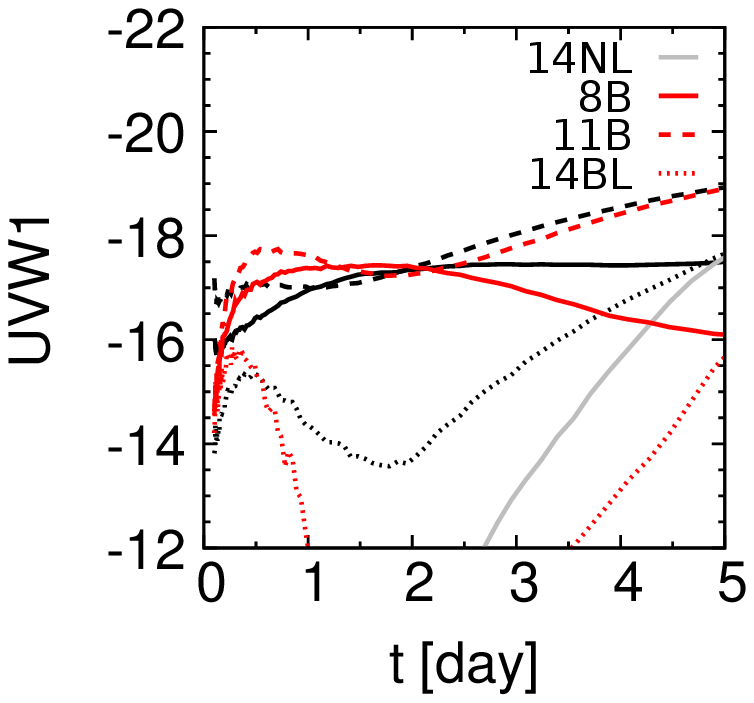}
        \end{minipage}\\
\hspace{-2cm}
        \begin{minipage}[]{0.45\textwidth}
                \epsscale{1.6}
                \plotone{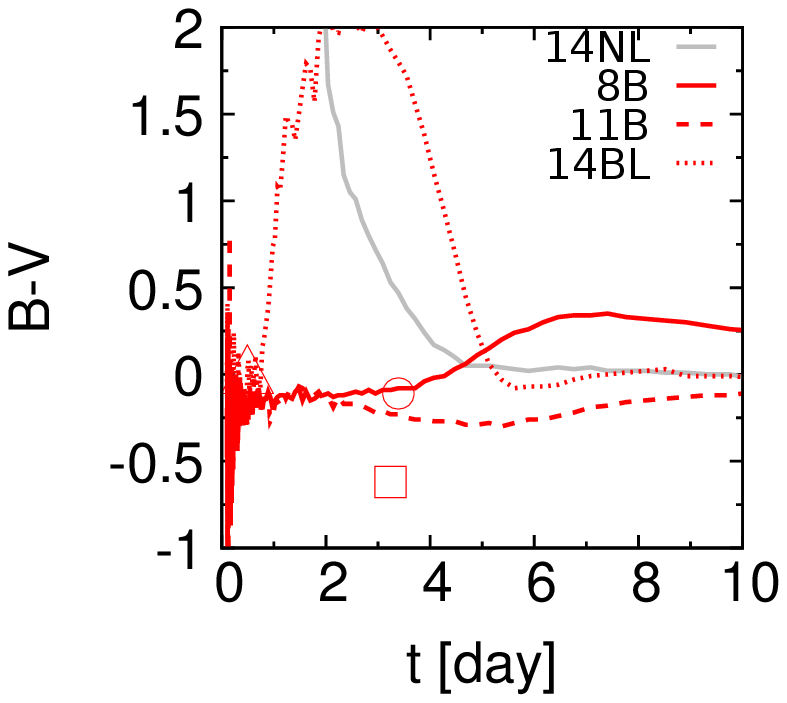}
        \end{minipage}
        \begin{minipage}[]{0.45\textwidth}
                \epsscale{1.6}
                \plotone{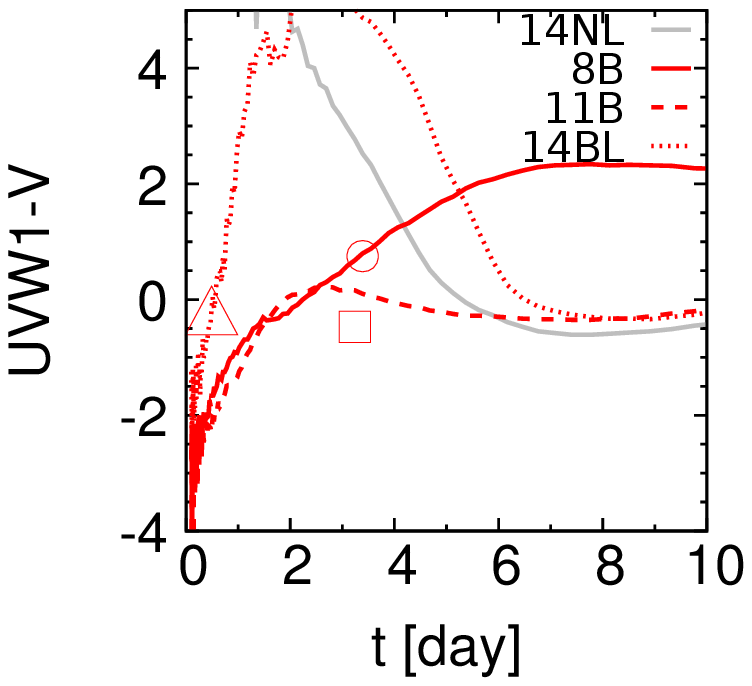}
        \end{minipage}
\end{center}
\caption
{The same as Figure 3 (He detonation models), but for the model sequence B for the detonation products. Note that the information on the $V$-band peak date for the 11B model as reconstructed from our `subtracted' light curves (shown by the open square) does not match to the fully synthesized emission properties as there is a significant contribution from the main SN ejecta in this model. The derived peak information is therefore highly uncertain for this particular model. 
\label{fig4}}
\end{figure*}

The additional source of the radiation from the inner $^{56}$Ni contribution may affect the light curves and color evolution especially for small binary separation. For $D = 5 \times 10^{11}$ cm, the expected bolometric luminosity from the $^{56}$Ni contribution may overwhelm the shock-thermalized radiation output. Given the expected slow color evolution of $^{56}$Ni contribution (similar to the He detonation scenario but with the higher temperature; \S 2.2 and \S 2.3.1), the optical output may especially be contaminated by this additional $^{56}$Ni contribution. On the other hand, the {\em UV} output may not be significantly contributed by this additional power source. This may affect the interpretation of the {\em UV}-optical behavior from the companion interaction scenario. Further investigation of this effect is beyond the scope of the present work, and will be presented elsewhere. 

\subsection{Radioactive decays of the He detonation products}

Figures 3 and 4 show the synthetic light curves of the He detonation models, for two different assumptions on the He detonation products (the sequence A and B; see Tables 1 \& 2 and \S 2.3.1). The figures show that two different choices on the He detonation products generally predict similar light curves. This can be understood by the similar energy input for the two sequences (\S 2.3). As such, we mainly discuss the results from the model sequence A in the following discussion. 

The bolometric light curves predicted for the He detonation scenario show different behavior than the CSM/companion interaction scenarios. Unlike the interaction scenarios, it shows the peak in the light curve in the {\em bolometric} luminosity. This is analogous to the usual $^{56}$Ni-powered main peak \citep{arnett1982,arnett1996}, as the photons require the diffusion time scale (more precisely, as combined with the dynamical time scale) to leak out of the system. This is longer for Model 8A/B than Models 11A/B and 14AL/BL, as our model sequence is constructed so that a more massive He layer is attached to a less massive progenitor WD. 

It is seen that the $V$-band light curves follow closely the bolometric light curves, especially visible for Models 8A/B and 14AL/BL\footnote{This behavior is not clear in Model 11A/B, since the radiation from the main SN ejecta already contributes significantly at the early-phase peak of the light curve powered by the surface He detonation products. As we constructed the model sequence following the double detonation scenario, a more massive WD is associated with a larger amount of $^{56}$Ni in the main ejecta with more extended distribution, leading to the fast rising in the main component (see also \S 6).}. Namely, there is not much bolometric correction unlike the CSM/companion interaction scenarios, and there is much less evolution in the characteristic temperature during the initial flash created by the He detonation products. Accordingly, the $B-V$ color stays at $\gsim 0$ during the phase, which will then evolve to the red when the rising light curve from the main SN ejecta starts dominating the total radiation output. The {\em UV}-optical color evolution shows a noticeable difference to the interaction models; the contrast in the color between the early flash and the later phase (e.g., $\sim 10$ days) is much smaller than the interaction models. 
Note that the {\em UV}-optical and $B-V$ colors of the He detonation model can first evolve even to the redder than the corresponding model without the He layer (i.e., seen by comparing the models 14AL/BL with Model 14NL). This characteristic behavior will be further discussed in \S 6. 

Detectability of the early flash by the decays of the He detonation products as a function of the WD mass depends on a combined effect, one by the luminosity and the time scale of the early flash and the other by the brightness of the underlying $^{56}$Ni-powered emission from the main SN ejecta. If we follow the relation in the WD mass and the He layer mass expected from the double detonation model sequence, the early flash by the He detonation products is more easily distinguishable for a less massive WD (due to the larger luminosity of the flash and the smaller luminosity from the main ejecta). If the properties of the main ejecta are fixed, then the early flash from the smaller amount of the He layer would be easier to distinguish, thanks to its shorter time scale (despite lower luminosity) which separates the early flash from the main body of the light curve (see Model 14AL/BL). 

\section{Predicted Behaviors of The Early `Flash' for Different Scenarios}

\begin{figure*}
\begin{center}
\hspace{-2cm}
        \begin{minipage}[]{0.45\textwidth}
                \epsscale{1.6}
                \plotone{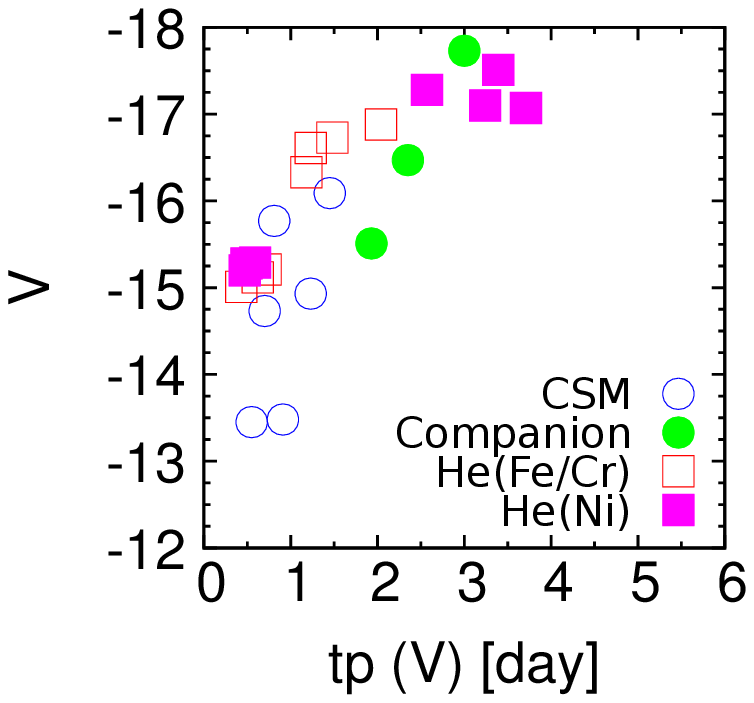}
        \end{minipage}
        \begin{minipage}[]{0.45\textwidth}
                \epsscale{1.6}
                \plotone{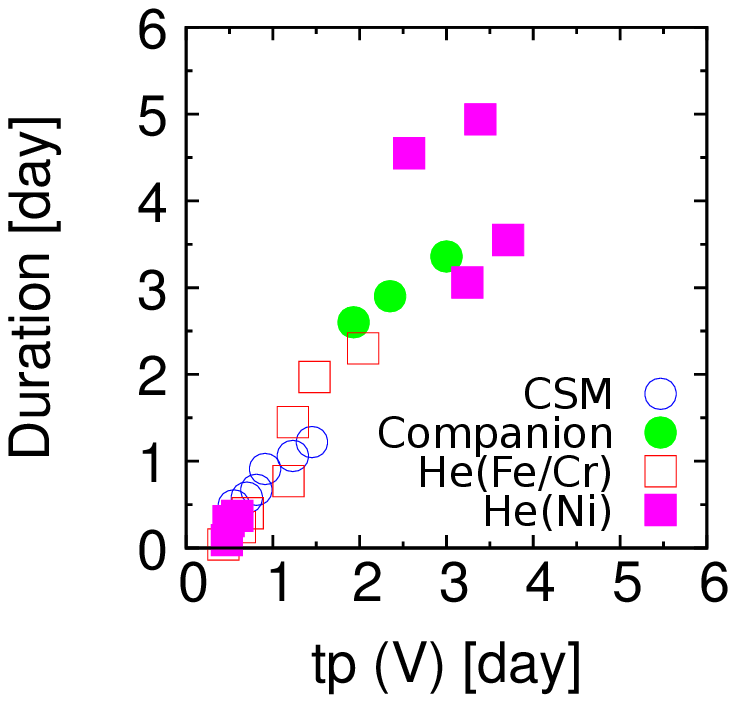}
        \end{minipage}\\
\hspace{-2cm}
        \begin{minipage}[]{0.45\textwidth}
                \epsscale{1.6}
                \plotone{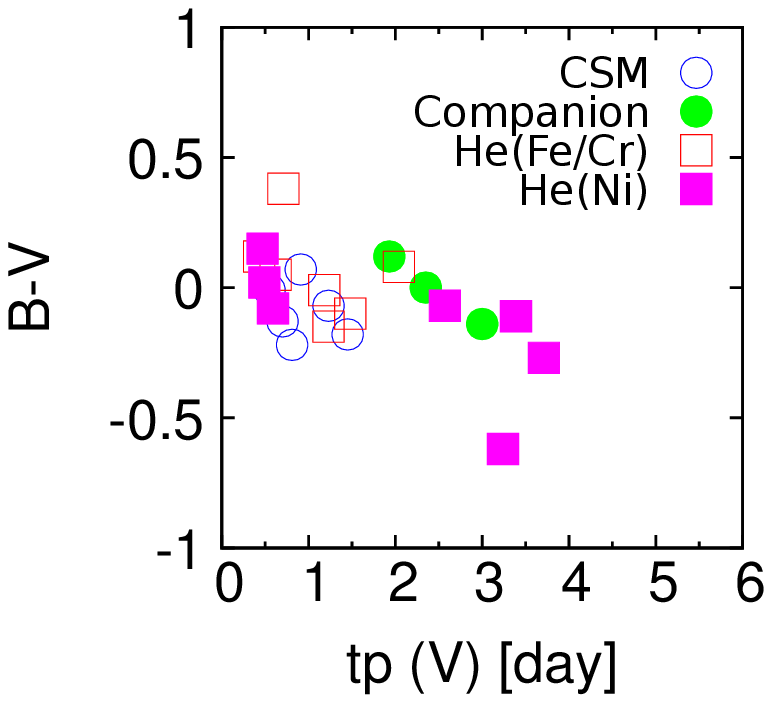}
        \end{minipage}
\end{center}
\caption
{Optical properties of the simulated models for the early excessive emission. Shown here are for the CSM interaction models (blue-open circles), the companion interaction models (green-filled circles), the He detonation models (the He detonation product sequence A and B by red-open and magenta-filled squares, respectively). Different panels show the peak $V$-band magnitude, duration of the $V$-band peak (defined as the duration where the magnitude is within $0.3$ mag below the peak), and the $B-V$ color at the $V$-band peak, as a function of the peak date in the $V$-band light curves. 
\label{fig5}}
\end{figure*}

\begin{figure*}
\begin{center}
\hspace{-2cm}
        \begin{minipage}[]{0.45\textwidth}
                \epsscale{1.6}
                \plotone{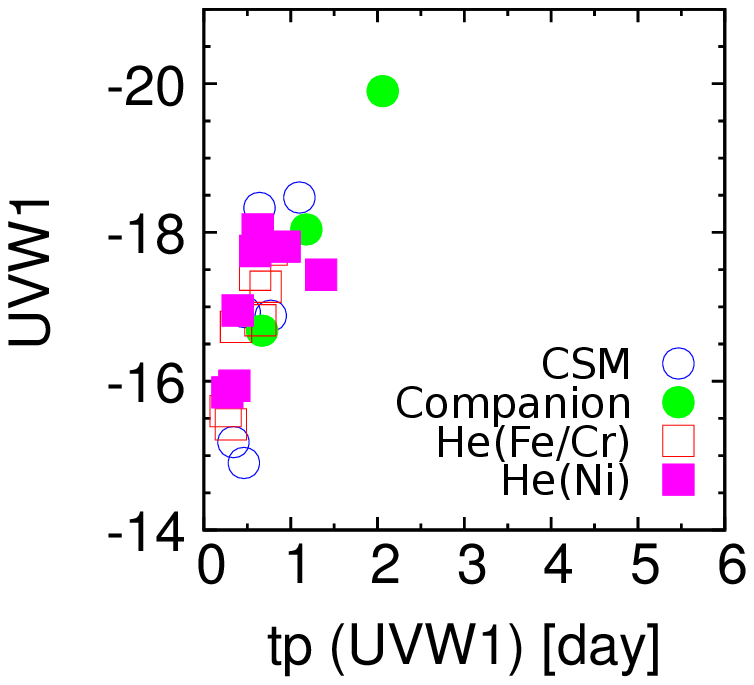}
        \end{minipage}
        \begin{minipage}[]{0.45\textwidth}
                \epsscale{1.6}
                \plotone{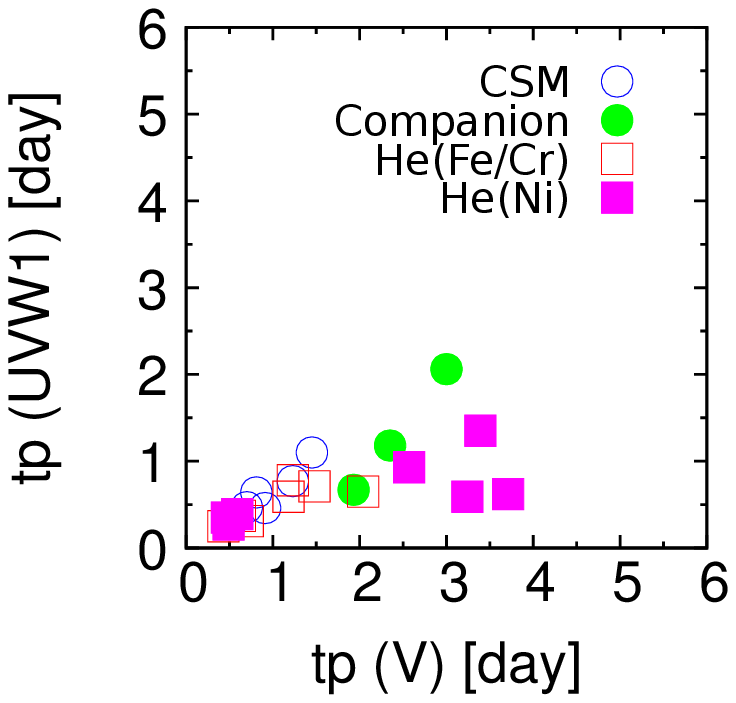}
        \end{minipage}\\
\hspace{-2cm}
        \begin{minipage}[]{0.45\textwidth}
                \epsscale{1.6}
                \plotone{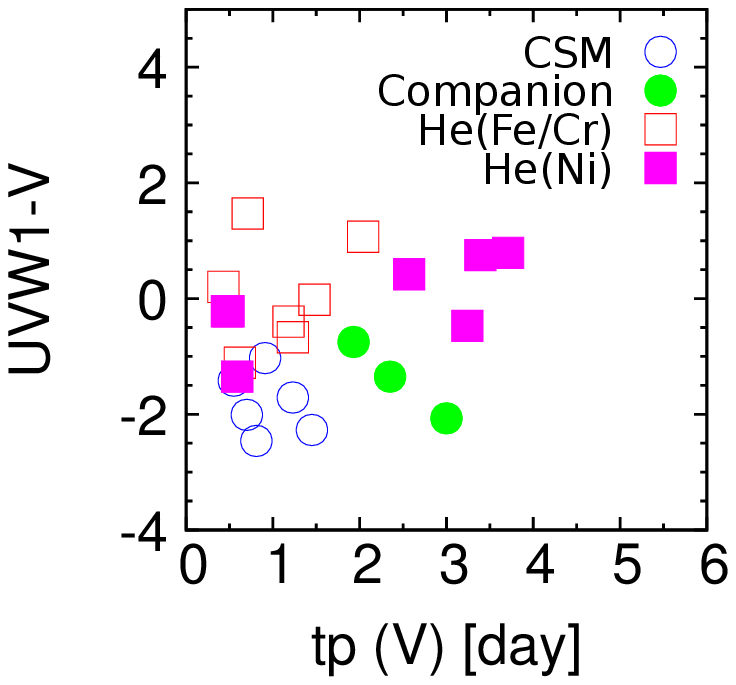}
        \end{minipage}
        \begin{minipage}[]{0.45\textwidth}
                \epsscale{1.6}
                \plotone{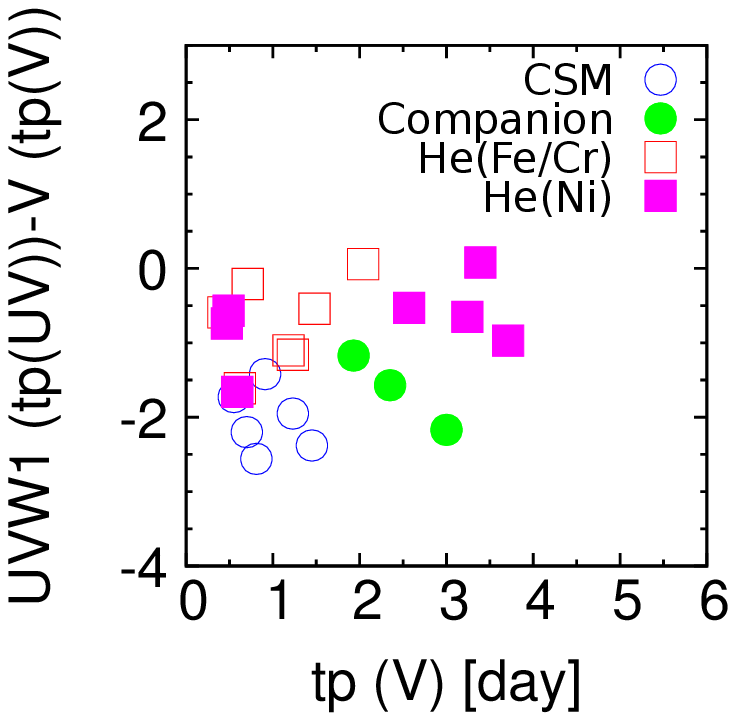}
        \end{minipage}
\end{center}
\caption
{{\em UV} properties of the simulated models for the early excessive emission. Shown here are for the CSM interaction models (blue-open circles), the companion interaction models (green-filled circles), the He detonation models (the He detonation product sequence A and B by red-open and magenta-filled squares, respectively). Different panels show the peak $UVW1$-band magnitude, duration of the $UVW1$-band peak (defined as the duration where the magnitude is within $0.3$ mag below the peak), the $UVW1$-$V$ color at the $V$-band peak, the color between the $UVW1$-band peak magnitude and the $V$-band peak magnitude (at each peak). The $UVW1$ peak magnitude is shown as a function of the $UVW1$-band peak date, while the other values are shown as a function of the $V$-band peak date.  
\label{fig6}}
\end{figure*}

At the first look, the predictions from different mechanisms may look quite similar and difficult to distinguish. However, characteristic behavior can indeed be different as different physical mechanisms are at work. This is especially the case between the model involving the interaction (either the CSM or companion) and that with the He detonation products. Figure 5 shows the characteristic features expected from different models in the optical. The features related to the {\em UV} emission are shown in Figure 6. 

For the He detonation models, the radiation transfer simulations are performed for the `whole' ejecta structure including the main part of the ejecta. We simulate the corresponding models without the He detonation ash on the surface (model sequence N), and the resulting light curves are subtracted from the corresponding models with the He layer, to obtain the early phase light curves from a `pure' He-detonation component. This could introduce a possible systematic error in deriving the characteristic observables, as the He detonation ash can also change the rising part of the main light curves by providing additional opacity source (see below). However, the early emission from the He detonation model is so distinct from the main SN component in many cases, and the underlying main SN component would not affect the extracted characteristic properties of the He detonation component in the first few days according to our visual inspection of the subtracted light curves. 

Figure 5 shows that there is a general trend that more delayed emission is associated with brighter peak luminosity. The SN-CSM interaction and companion interaction scenarios share the similar relation, as these two scenarios involve the same physical mechanism to create the early emission, i.e., the cooling emission where the optical peak is created by the decreasing temperature associated with monotonically decreasing bolometric luminosity.

While predicting a qualitatively similar relation, the relation seen in the He detonation model sequence is originated from a totally different reason. A larger amount of the He layer leads to both the large amount of the energy source (thus larger luminosity) and a larger amount of the opacity source (thus longer diffusion time scale). On average, the He detonation scenario predicts the $V$-band peak magnitude brighter than the CSM/companion interaction scenarios by $\sim 1$ magnitude for a given peak date, but the quantitative comparison is subject to the uncertainties in constructing the (simplified) interaction models. 

The duration of the early phase `flash' as a function of the peak date is generally similar between the interaction scenarios and the He detonation scenario. The duration here is defined as the time period during which the luminosity stays within 0.3 mag below the (early-phase flash) peak. There is a general trend that the duration is longer for more delayed peak emission.

\begin{figure*}
\begin{center}
\hspace{-2cm}
        \begin{minipage}[]{0.45\textwidth}
                \epsscale{1.6}
                \plotone{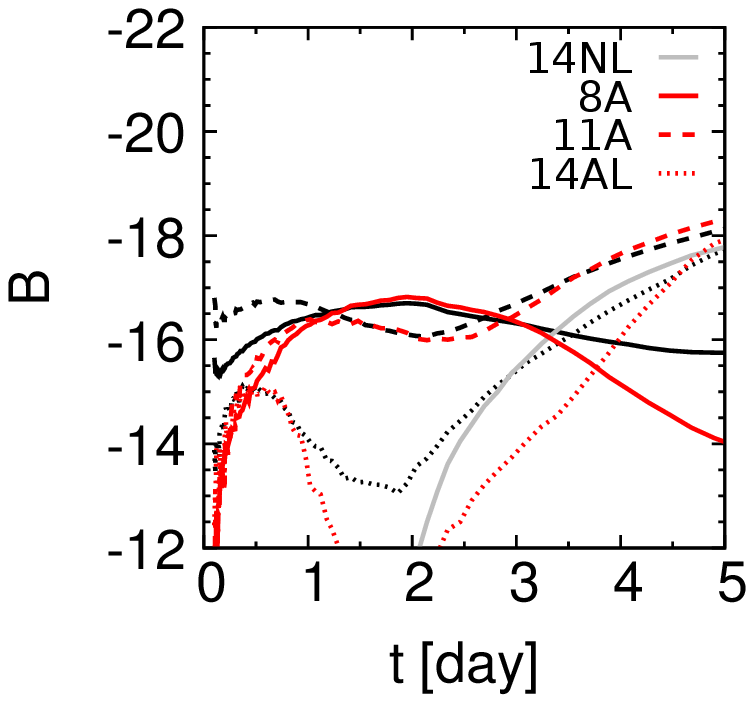}
        \end{minipage}
        \begin{minipage}[]{0.45\textwidth}
                \epsscale{1.6}
                \plotone{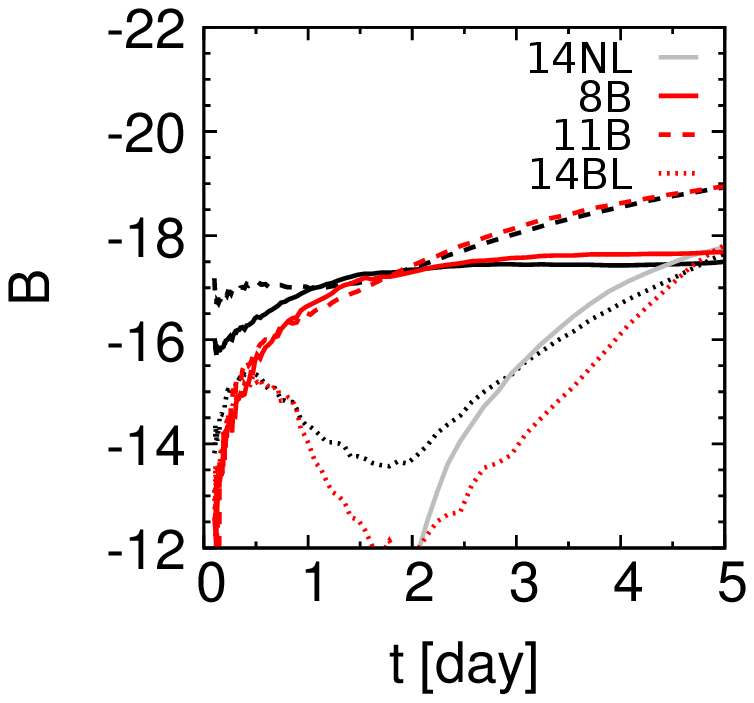}
        \end{minipage}
\end{center}
\caption
{$B$-band light curves of the He detonation models. See the caption of Figure 3 for different line styles. 
\label{fig7}}
\end{figure*}

The $B-V$ colors at the peak are also shown in Figure 5. In general, the He detonation model sequence predicts the peak $B-V$ color of $\sim 0$, similar to the interaction models. This behavior is indeed not trivial; the (optical) peak in the interaction models is a result of the characteristic behavior in which the optical peak is created once the SED peak moves to the optical wavelength in terms of the cooling emission, and the color is set by this requirement. This condition does not necessarily apply to the He detonation scenario, which shows both the bolometric and optical luminosities can increase. Figure 7 shows the $B$-band light curves of the He detonation model sequence. This figure shows that $B$-band light curves generally follow the bolometric light curves, similar to the case for the $V$-band light curves. Namely, the bolometric correction is small and it does not evolve significantly, 

\begin{figure*}
\begin{center}
\hspace{-2cm}
        \begin{minipage}[]{0.32\textwidth}
                \epsscale{1.6}
                \plotone{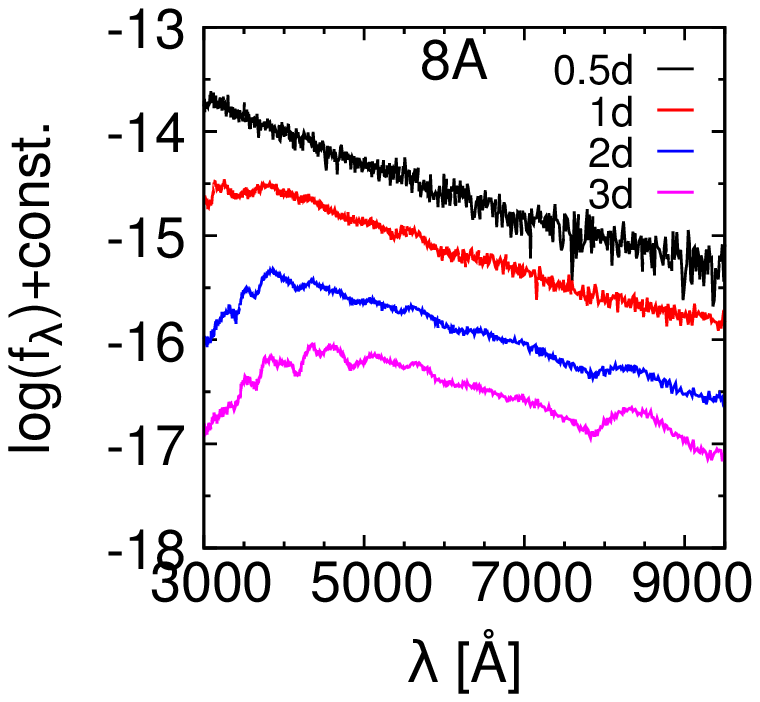}
        \end{minipage}
        \begin{minipage}[]{0.32\textwidth}
                \epsscale{1.6}
                \plotone{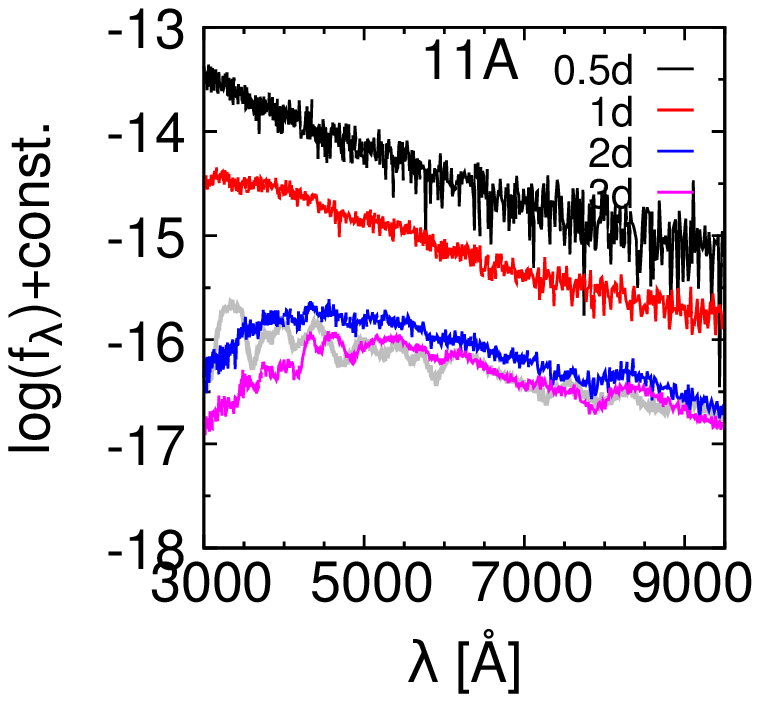}
        \end{minipage}
        \begin{minipage}[]{0.32\textwidth}
                \epsscale{1.6}
                \plotone{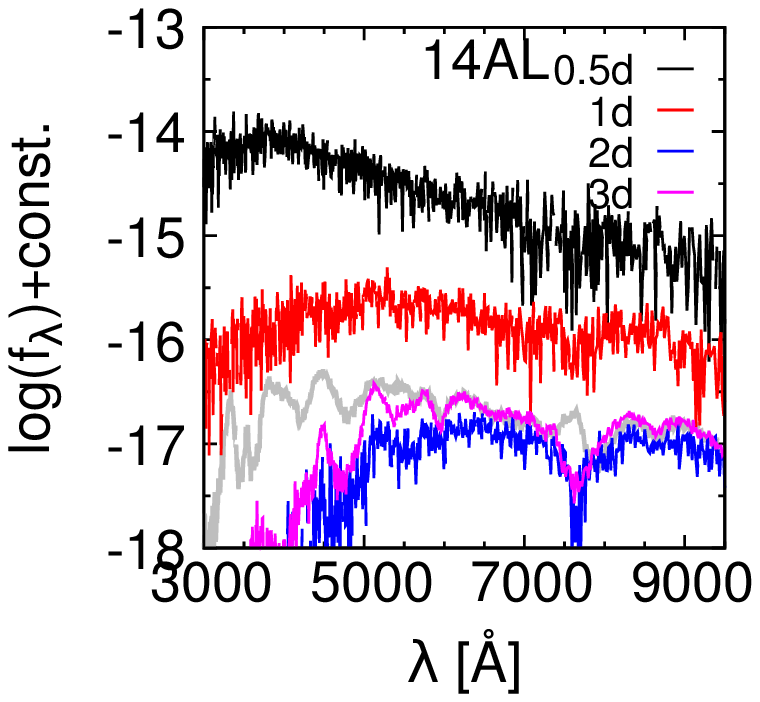}
        \end{minipage}
\end{center}
\caption
{Spectra of the He detonation models (the He detonation product sequence A) in the first few days after the explosion. The spectra of the corresponding models without the He detonation products at $3$ days is also shown (gray line; 11N and 14NL corresponding to 11A and 14AL). 
\label{fig8}}
\end{figure*}

From the above investigations, we conclude that the He detonation scenario may be difficult to distinguish from the companion interaction scenario only with the optical photometric information in the early flash. The early phase behavior is indeed more easily distinguishable once the information in the {\em UV} wavelength is combined. Figure 6 shows that there is a general trend between the {\em UV} peak date and the peak luminosity, irrespective of the mechanisms, where the brighter peak is reached for more delayed emission. The difference in the peak {\em UV} magnitude between the He detonation scenario and the CSM/companion interaction scenarios is indeed not large for a given {\em UV} peak date.  However, the He detonation scenario predicts that the {\em UV} peak is reached quite early, and the time lag between the {\em UV} peak and the optical peak tends to be larger for the He detonation model than the others. This stems from the cooler photosphere in the He detonation model especially in the first few days (i.e., increasing bolometric luminosity in the He detonation model as compared to the decreasing bolometric luminosity in the interaction scenarios). For the same reason, the {\em UV}-optical color is much redder in the He detonation scenario in the first few days. Once the optical and {\em UV} behaviors are combined, the He detonation scenario thus behaves in a different manner than the interaction scenarios. The {\em UV}-optical color is much redder in the He detonation model sequence than the CSM/companion interaction scenarios (by $\sim 1-2$ mags), as shown by Figure 6 both for the $UVW1$-$V$ color at the $V$-band peak and for the difference between the peak $UVW1$-band magnitude and the peak $V$-band magnitude.

\section{Spectral Signatures in the He detonation Flash and effects of absorptions}

Figure 8 shows synthetic spectral evolution in the first 3 days for selected He detonation models, 8A, 11A and 14AL for which the light curves are available in Figures 3 and 7. In the optical, these models peak at $\sim 2$, $1$, $0.5$ days since the explosion, respectively for models 8A, 11A and 14AL. It is seen that the optical spectra show blue continuum before the peak date. Around the peak, the line features start developing, mostly by Fe and Fe-peak elements but Ca II absorptions are also strong. These absorptions are substantially contributed by the He detonation products. For models 11A and 14AL, the emission from the `main' SN ejecta indeed contributes predominantly at $\sim 3$ days. The comparison between models 11A/14AL with the corresponding models 11N/14NL (without the He layer) shows that stronger absorptions are associated with the He detonation models. It is, however, not determined merely by the amount of the He detonation products. Model 14AL indeed has a smaller amount of the He ash (nearly by an order of magnitude) but shows stronger absorption. The bolometric magnitude is $\sim -17$ mag for 11A but $\sim -15$ mag for 14AL, at $\sim 3$ days. This difference in the luminosities results in the difference in the photospheric/characteristic temperature roughly by $\gsim 50$\% (assuming that the photospheric radii are similar). This large difference can make difference in the ionization status, and the lower temperature in 14AL results in the stronger absorption (due to the enhanced abundance of singly-ionized ions). 

Indeed, this effect of the absorption by the He detonation ash is already seen in the $V$-band (Figure 3) and $B$-band (Figure 7) light curves. In the $V$-band, models 14AL and 14N show very similar light curves in the rising part of the main SN component (after $\sim 2$ days). On the other hand, in the $B$-band, model 14AL is fainter than 14NL by $\sim 1$ magnitude at $\sim 3$ days. The difference between models 14AL and 14NL becomes smaller as the luminosity rises, and the two models overlap at $\sim 5$ days when the bolometric luminosity reaches to $\sim -18$ mag. 

This additional absorption is also key in the characteristic color evolution in the He detonation scenario. Not only the relatively red (early) peak color, but also subsequence evolution can be distinguished for the He detonation scenario. For the CSM and companion interaction scenarios, one expects that after the main SN ejecta start dominating the emission, the color follows this component as well. For example, Figure 1 shows that the CSM interaction model initially shows blue emission with $B-V \lsim 0$. The color quickly evolves to the red, and the early-phase $V$-band peak is reached when $B-V \sim 0$. The emission from this cooling emission is further reddened, but then this reddening is saturated by the SN ejecta component, which will subsequently become blue and reaches back to $B-V \sim 0$ after a few days (see model 14NL in Figure 1). While we have not performed the radiation transfer including both components, this behavior is expected since there is no additional source of opacity and is indeed seen in numerical simulations by \citet{piro2016}. The similar behavior is expected for the companion interaction as well, and in this case due to the longer time scale of the initial flash it will not show the `red phase'; it is initially very blue, and then in a few days merges to the main SN emission at $B-V \sim 0$. 

This can be different in the He detonation scenario. Figure 3 shows that model 14AL becomes redder than model 14NL until $\sim 6$ days. During this phase, the luminosity of the main SN component is still small, resulting in low temperature in the outermost layer occupied by the He detonation ash. Therefore, additional absorption is created by singly-ionized Ca, Fe and Fe-peaks. This effect is relatively minor for model 11A, due to the large luminosity and early emergence of the SN main component, which keep the high ionization in the He layer. Model 8A becomes very red, as the main SN component never reaches $\sim -18$ mag ($M$($^{56}$Ni) $= 0.17 M_\odot$ in the main SN ejecta) and thus the He layer is always kept at low ionization level (see below for further details). 

The (relatively) blue continuum and shallow absorptions may qualitatively share features expected for the CSM/companion interaction scenarios, while the He detonation scenario tends to produce redder color. The quick reddening after the peak of the early flash could also be found in all the scenarios. The development of the strong absorption features in the post-flash phase, correlated with the luminosity, can be a characteristic feature of the He detonation scenario. Further quantitative comparison between the He detonation scenario and the interaction scenarios on the last point will require detailed radiation transfer simulations for the CSM/companion interaction scenarios, which we postpone to the future. Still, it is expected that the CSM/companion interaction scenarios eventually converge to the non-interacting models (such as the model sequence N), while the He detonation scenario does provide additional source of the absorption, and thus the difference is expected. 

\section{The Maximum-Light Behaviors}

\begin{figure*}
\begin{center}
\hspace{-2cm}
        \begin{minipage}[]{0.32\textwidth}
                \epsscale{1.6}
                \plotone{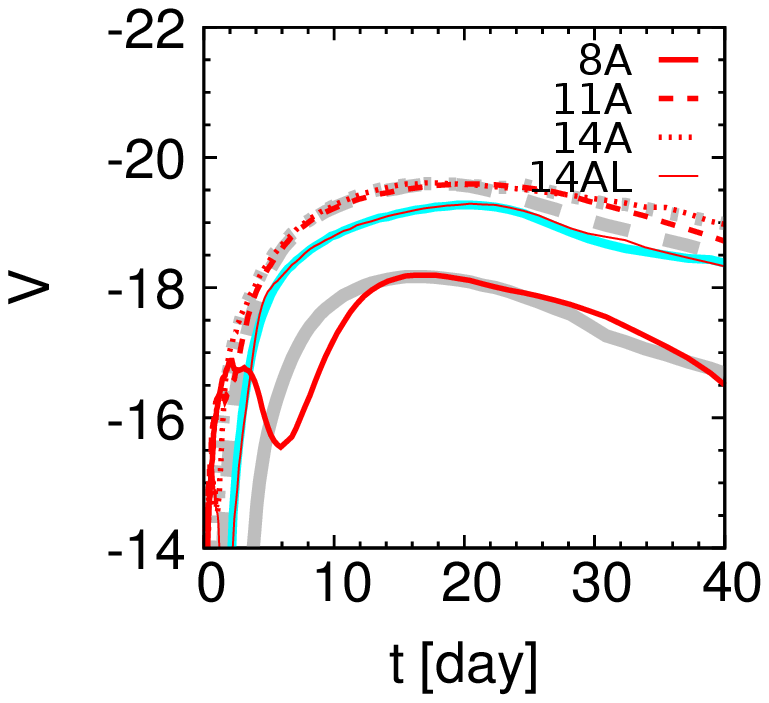}
        \end{minipage}
        \begin{minipage}[]{0.32\textwidth}
                \epsscale{1.6}
                \plotone{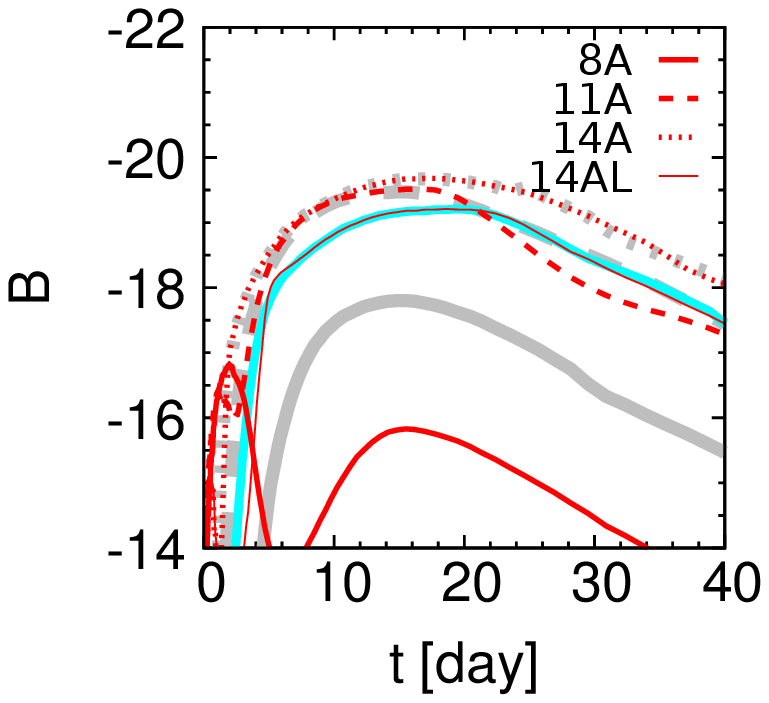}
        \end{minipage}
        \begin{minipage}[]{0.32\textwidth}
                \epsscale{1.6}
                \plotone{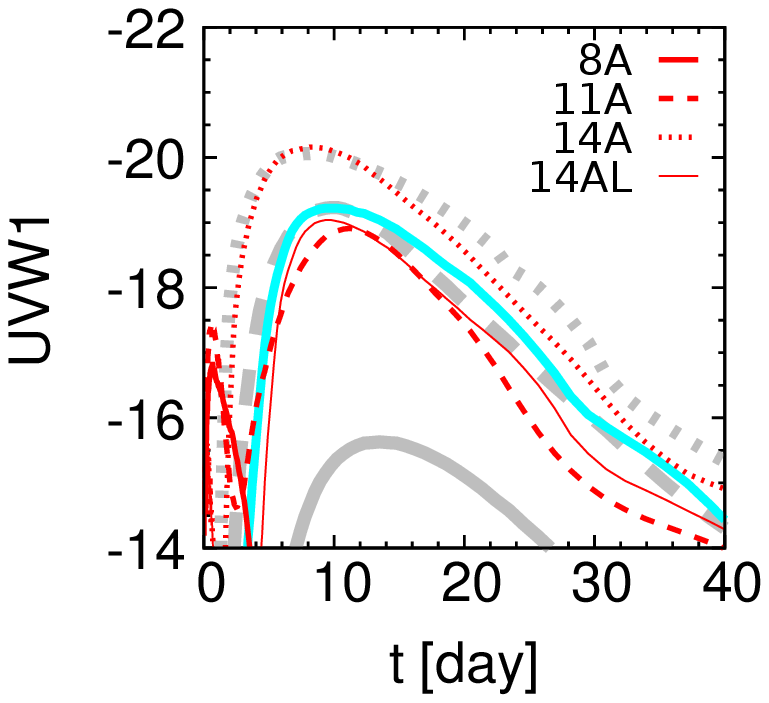}
        \end{minipage}
\end{center}
\caption
{Light curves of the He detonation models covering the maximum light. Shown here are models 8A (red thick-solid), 11A (red dashed), 14A (red dotted) and 14AL (red thin-solid). The corresponding models without the He detonation products are also shown by gray (8N, 11N, 14N) and sky blue (14NL) lines with the same line styles. Different panels show the $V$-, $B$-, and $UVW1$-band light curves. 
\label{fig9}}
\end{figure*}

Maximum light behaviors expected for the double detonation scenario have been discussed by several groups mostly in the optical \citep{kromer2010,woosley2011}. Here, we briefly address the issue especially in relation to the behavior in the very early phase, also adding discussion in the {\em UV}. Figures 9 and 10 show $V$, $B$, and $UVW1$ light curves and spectra around the maximum light ($10$ to $30$ days since the explosion) for a few selected models. 

Figure 9 shows that for the `bright' models (reaching to $\sim -19$ mag or brighter), the He detonation ash does not substantially change the optical light curves up to $\sim 40$ days. In our model sequence (except for 14AL/BL), a more massive WD is associated simultaneously with the larger amount of $^{56}$Ni in the main SN ejecta and the smaller amount of the He layer. Therefore, the models with larger luminosity is generally associated with the smaller amount of the additional opacity source (obviously, the light curves should converge to the model without He layer for the decreasing amount of the He layer). Furthermore, larger luminosity keeps the temperature high, preventing the increase of the opacity provided by singly-ionized Ca, Fe, and Fe-peaks in the He detonation layer. 

The optical light curves of Model 14A in the maximum phase are basically identical to those of Model 14N, as the absorption in the He layer is negligible. This also applies to model 14AL, which has the same mass of the He layer with 14A but with smaller luminosity. The importance of the mass of the He layer is seen by comparing Models 14AL and 11A. The peak (bolometric) luminosity is indeed higher in model 11A (with the larger amount of $^{56}$Ni) than Model 14AL, while it starts showing additional absorption in the $B$-band in the post maximum following the luminosity decrease. Note that the mass of the He layer is larger in 11A than 14AL by an order of magnitude. For Model 8A, this additional absorption by the He detonation products is very strong. This has both the large amount of the He layer ($\sim 0.1 M_{\odot}$) and the small amount of $^{56}$Ni in the main SN ejecta ($\lsim 0.17 M_{\odot}$). For Model 8A, even the $V$-band initially shows the suppression in the rising phase as compared to Model 8N. This is due to the formation of the photosphere at high velocity due to the large opacity, leading to the low temperature (see below for the spectra in this phase). 

The effect is stronger in the {\em UV} because of the substantial line blending. Still, for models peaking at the magnitude brighter than $\sim -19$ mag, the effect in the maximum and post-maximum is not substantial. The effect is stronger in the rising part (see Fig. 3), because of the combination of the low luminosity and the high density. The sub-luminous models (e.g., Model 8A) are predicted to be fully blacked out in the {\em UV}, highlighting the {\em UV} behavior as a distinguished diagnostic of the He detonation scenario. 

\begin{figure*}
\begin{center}
\hspace{-2cm}
        \begin{minipage}[]{0.32\textwidth}
                \epsscale{1.6}
                \plotone{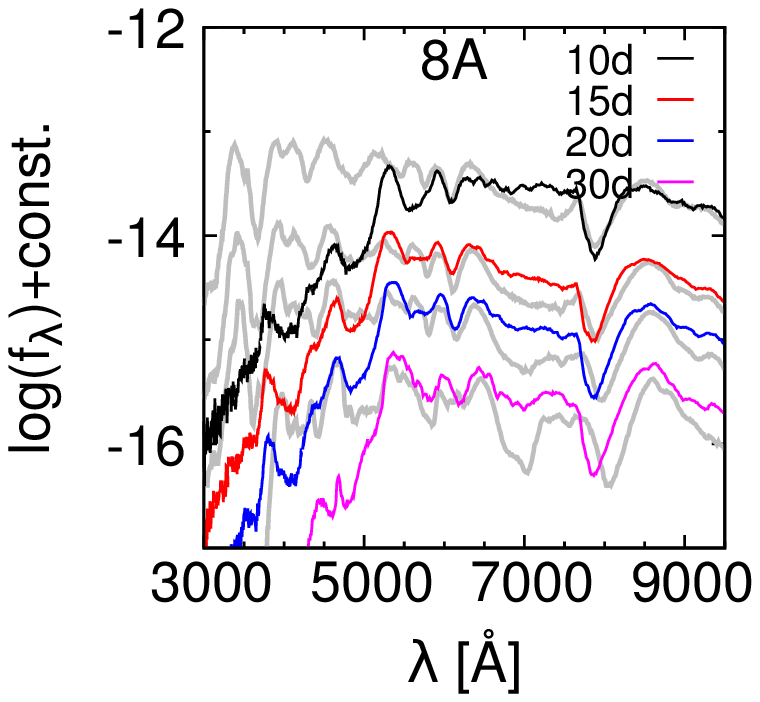}
        \end{minipage}
        \begin{minipage}[]{0.32\textwidth}
                \epsscale{1.6}
                \plotone{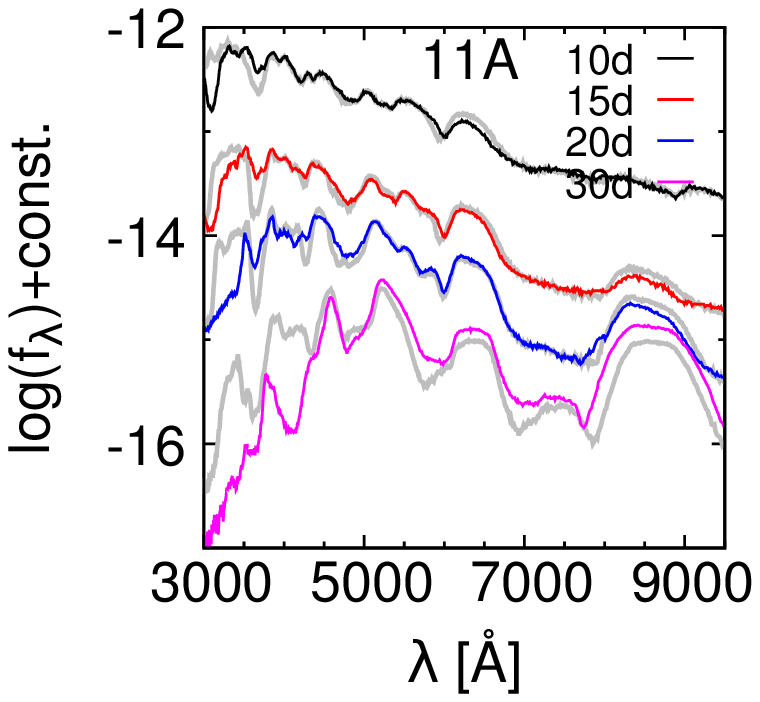}
        \end{minipage}
        \begin{minipage}[]{0.32\textwidth}
                \epsscale{1.6}
                \plotone{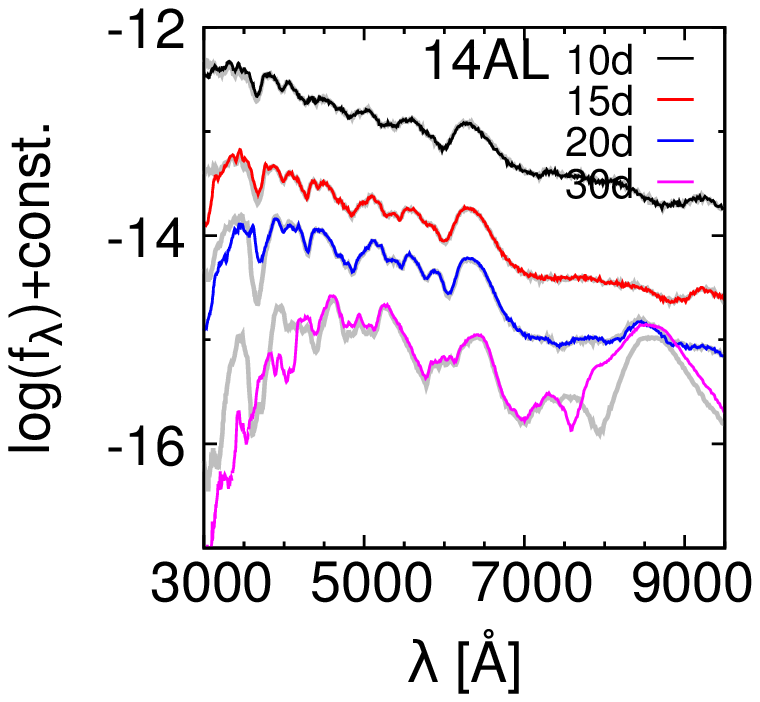}
        \end{minipage}
\end{center}
\caption
{Spectra of the He detonation models (the He detonation product sequence A) around maximum light. The spectra of the corresponding models without the He detonation products at the same epochs are also shown (gray lines; 8N, 11N, 14NL corresponding to 8A, 11A, and 14AL). 
\label{fig10}}
\end{figure*}

Figure 10 shows the spectral evolution from 10 to 30 days since the explosion. To see the effect of the He detonation ash, the spectra are overplotted with the corresponding models without the He layer. For Model 14AL, only the spectra in the post-maximum is affected by the He layer. There are additional absorptions in the blue, e.g., Ti trough at $\sim 4,000$\AA,  which start developing in the post-maximum. The behavior is similar for model 11A, but the effect becomes in active earlier; already around the maximum there is a hint of the additional absorption, and in the post maximum the blue portion of the spectrum is substantially absorbed with clear Ti trough. The situation is dramatically different for Model 8A. Already in the rising phase the spectrum is very red. Overall the spectral signature shares similarities to sub-luminous 1991bg-like SNe Ia, while the color is even redder than 1991bg-like SNe, lacking totally the UV emission. These behaviors can be understood by a combination of the mass of the He detonation ash and the luminosity provided by the main SN ejecta, as already discussed throughout the paper. 

\section{Implications for SNe Ia with possible early flash}

There are a few SNe Ia so far for which possible signatures of the early excessive emission in the first few days have been reported. \citet{cao2015} reported the initial {\em UV} excess for iPTF14atg. iPTF14atg is a peculiar SN Ia which shares similarities to SN  2002es (sub-luminous SNe Ia with slow decay). They found that the initial {\em UV} luminosity and color are well explained by the interaction with a non-degenerate companion star according to the model by \citet{kasen2010}, most likely with a radius of $\sim 60 R_{\odot}$. 

\citet{marion2016} reported blue color for SN 2012cg up to $\sim -15$ day (measured from the $B$-band maximum), which is further complemented with a possible excess in the {\em UV} at the beginning. By comparing the optical light curves and color evolution with the companion interaction model, they suggested that this could be a result of the interaction with a non-degenerate companion star, most likely a MS of $\sim 6 M_{\odot}$ ($D \sim 2 \times 10^{12}$ cm assuming the Roche lobe overflow). SN 2012cg shares similarities to SN 1999aa, which is slightly over-luminous and slow decliner, with the light curve and spectral features generally between normal and 1991T-like SNe Ia. 

\citet{hoss2017} presented extensive data set for SN 2017cbv. Spectroscopically it resembles normal SN Ia 2013dy. Its declining rate ($\Delta m_{15}$ ($B$)) would also indicate its normal classification. Contrarily, the estimated luminosity is high ($M_{\rm B} = -20.04$ mag) which however is a subject of the uncertainty in the poorly constrained distance to the host galaxy while the extinction would not be substantial. Its relation to SN 2012cg is not clear; the optical light curves and color evolution are similar between the two at $\gsim$ one week after the explosion, while SN 2017cbv seems to be much bluer in the initial $\sim 5$ days where the excessive emission is found (see below for further details). \cite{hoss2017} discussed several scenarios, including the companion interaction, CSM interaction, and the contribution by radioactive isotopes near the surface which may originate either from extensive $^{56}$Ni mixing or the He detonation. 

Recently, \citet{jiang2017} reported the discovery of the very early detection of an SN Ia, named MUSSES1604D. At the discovery the $g$-band magnitude was $\sim -14$ mag, and in one day it was brightened to $\sim -16$ mag in the $g$- and $r$-bands. Then the $g$-band magnitude stayed at the similar magnitude, suggesting that the discovery was within a day after the explosion and the early flash was peaked at $\sim -16$ mag at 1-2 days after the explosion. The color evolved quickly from $B-V \sim 0$ to redder. Additionally, the follow-up observations show that MUSSES1604D was relatively normal in its peak luminosity ($V \sim -19$ mag) as is roughly consistent with its light curve declining time scale. The color around the maximum light was relatively red ($B-V \sim 0.3$ mag). It also shows interesting peculiarity in the maximum-light spectra, which show strong absorptions in the blue together with clear Ti trough at $\sim 4,000$\AA\ as typically seen in SN 1991bg-like sub-luminous objects but otherwise show similarity to normal SN Ia spectra. 

In this section, we compare our models with these examples and discuss possible origin(s) of the reported early flashes. We do not aim at providing detailed models tuned to fit to the observational data of these SNe in this paper. Rather, we try to clarify the applicability of the different scenarios based on the expectations by considering the physical processes involved in a qualitative way, in order to provide a guide for future detailed modeling and for future observations of similar events (or SNe Ia soon after the explosion, in general). The discussion on the maximum-phase light curves should be taken as being only qualitative as this can be affected by the detailed distribution of $^{56}$Ni and opacity sources in the main ejecta; indeed, our reference model 14NL shows relatively rapid rise as compared to normal SNe Ia. We further note that our companion interaction model prediction relies on the hypothesis of the equipartition between the thermal (gas) energy and the photon energy. Applicability of this assumption is beyond the scope of this paper \citep[e.g.,][]{kutsuna2015}. An additional caveat is that our prediction for the companion interaction is limited to the direction to the companion star. In the intermediate angles we would expect that the luminosity would become fainter than our prediction, while the color would not be highly dependent on the viewing direction.

\subsection{MUSSES1604D}

The nature of MUSSES1604D was discussed by \citet{jiang2017}. They concluded that the SN was triggered by the He detonation on the surface. In this section we briefly summarize their arguments to set a scene for the following discussion for the other objects. We refer \citet{jiang2017} for full details of the interpretation and arguments for the He detonation origin. 

The combination of the time scale ($\sim 1-2$ days) and the luminosity ($\sim - 16$ mag in the optical) observed for the early flash is inconsistent with the companion interaction scenario (Fig. 5). The $B-V$ color of the early flash is also predicted to be too blue with $B-V \lsim 0$ for the CSM and companion interaction scenarios. 

The He detonation provides the most consistent and straightforward interpretation. The observed time scale and brightness of the flash are along the expectation (Fig. 5) if the He layer is as massive as $\sim 0.03 M_\odot$. The relatively red color at the peak of the flash is also consistent. Additionally, the mass of the He detonation ash as massive as $\sim 0.03 M_\odot$ as combined with the normal maximum-light luminosity is accompanied by the significant, but not too strong, absorptions in the blue around the maximum light, including Ca II H \& K and the Ti trough. Therefore, the main features of the early flash and the maximum-light behavior including the spectral features can be naturally explained by the He detonation model consistently. 

The model shown in \citet{jiang2017} provides a reasonably good fit to the multi-band light curves (including the first few days and the maximum phase) and the maximum spectrum of MUSSES1604D. The model is largely similar to model 14AL in the main ejecta structure (but with the smaller mass of $^{56}$Ni, $\sim 0.4 M_{\odot}$), while the mass of the He layer is $\sim 0.02 M_\odot$ as is similar to model 11A. Due to the combination of the massive He layer and relatively low luminosity from the main SN ejecta, the model shows substantial absorption of the Ti and Fe peaks in the blue, which provides a good fit to the maximum-phase spectrum. Note that the mass of the $^{56}$Ni derived for MUSSED1604D is smaller than both models 14AL and 14A, thus the effect of absorption for MUSSES1604D is more significant than in 11A.This highlights the importance of not only the mass of the He layer but also the luminosity in shaping the spectra through the He detonation layer. Also, this suggests that potentially a large number of the He detonation-triggered explosions may be hidden; similar SN Ia explosions but with a slightly larger amount of $^{56}$Ni than MUSSES1604D may not show the characteristic absorption by the He detonation ash, and then it may be classified as a normal SN.

An unresolved problem would remain in the properties of the SN main component, between the ejecta mass and the luminosity (thus $^{56}$Ni in the main SN ejecta). The evolution of the light curve around the maximum-light indicates the ejecta mass similar to that in normal SNe Ia, favorably near the Chandrasekhar mass. The peak magnitude is relatively normal. This feature is not readily explained by the existing model sequence of the He detonation-triggered explosion. If the carbon detonation at the center (as induced by the shock wave propagation created originally by the He detonation on the surface) is responsible for the whole disruption of the WD, we expect a relation between the progenitor WD mass and resulting $^{56}$Ni, as a more massive and dense WD produces a larger amount of $^{56}$Ni. This behavior is seen in our model sequence (Table 1) which is constructed following the simulation results of the double detonation models (except for Models 14AL/BL/NL). In this scenario, SNe Ia with normal luminosity is predicted to be followed by fast decline. The same relation should also apply to the He-ignited violent merger scenario (see footnote 4). Understanding the exact nature of the physical mechanism leading to the phenomenologically derived properties of the He layer and the main SN component will require further investigations. 

\subsection{iPTF14atg}

\begin{figure*}
\begin{center}
\hspace{-2cm}
        \begin{minipage}[]{0.45\textwidth}
                \epsscale{1.6}
                \plotone{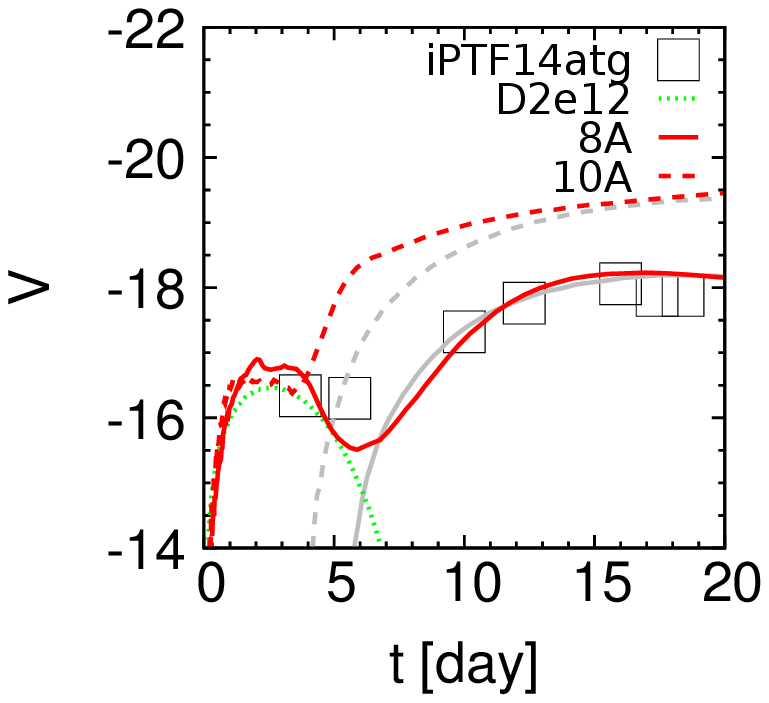}
        \end{minipage}
        \begin{minipage}[]{0.45\textwidth}
                \epsscale{1.6}
                \plotone{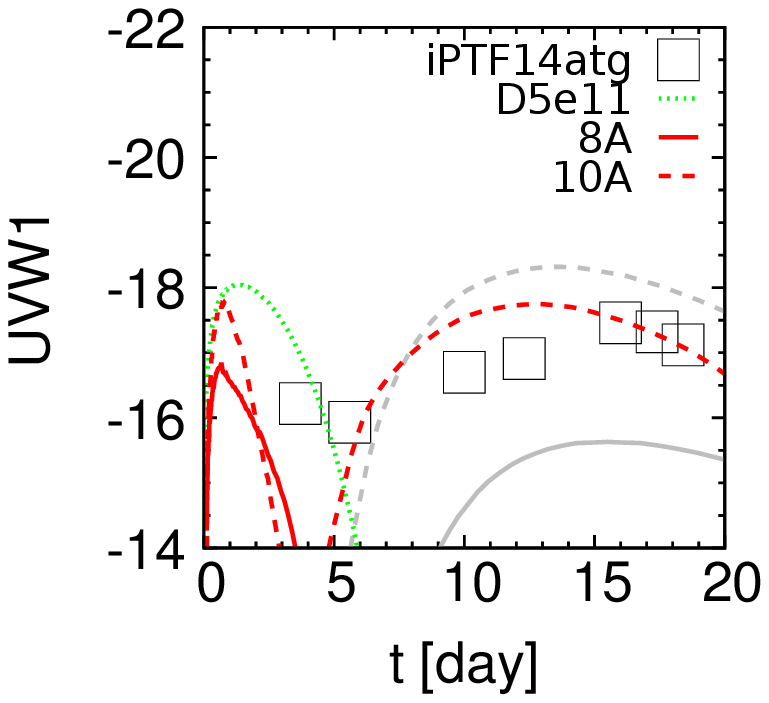}
        \end{minipage}
\end{center}
\caption
{The $V$-band and $UVW1$-band light curves of iPTF14atg \citep[open squares;][]{cao2015}, as compared to a few selected models. Shown here are the companion interaction model (with the separation of $2 \times 10^{12}$ cm; green dotted), the He detonation models 8A (red solid) and 10A (red dashed). The corresponding models without the surface He detonation products, 8N and 10N, are also shown (gray, with the line styles following the corresponding models 8A and 10A). 
\label{fig11}}
\end{figure*}

The $V$-band and $UVW1$-band light curves of iPTF14atg are shown in Figure 11. We apply the distance modulus of 34.9 mag. We assume negligible extinction even in the $UV$ \citep{cao2015}. The absolute magnitudes are computed using the {\em SWIFT UVOT} Vega system zero-points \citep{breeveld2010}. We adopt the explosion date as suggested by \citet{cao2015}. 

We first compare the observationally derived characteristic behaviors to the properties expected from each model (Figures 5 and 6). The $V$-band peak in the early flash might have been reached at $\sim 4$ days, and the $UV$ peak must have occurred before that. Therefore, the constraint is that the $V$-band peak magnitude must be brighter than $\sim -16$ mag within $\sim 4$ days. This leaves a subset of the companion interaction and the He detonation models as possible candidates. The $UVW1$-band peak magnitude must be brighter than $\sim -16$ mag, which is satisfied by basically all the models. Finally, the flat evolution in the $V$-band suggests that the $V$-band peak was reached around the first point, and the $UVW1 - V$ color at that moment is about $\sim 0$ mag or bluer. Since there is a sign of the luminosity decrease in the {\em UV} between the first and second points, it indicates that the $UVW1 - V$ pseudo-color measured at each peak is bluer than $\sim 0$ mag. These additional constraints are also satisfied by some of the models in the companion interaction and the He detonation scenarios. 

Therefore, the characteristic behaviors are explained both by the companion interaction scenario and the He detonation scenario. In the context of the companion interaction scenario, a rough match is found for the separation of $\sim 2 \times 10^{12}$ cm, as shown in Figure 11. In this scenario, the {\em UV} luminosity must have been much higher if the observation would have been performed earlier, highlighting the importance of the quick {\em UV} follow-up (or even high cadence surveys in the {\em UV}) to identify the origin of the early flash. 

In the He detonation scenario, the mass of the He layer should be $\gsim 0.02 M_{\odot}$. The exact criterion will depend on the details of the He detonation nucleosynthesis, but it will roughly be a correct order as both the sequences A and B require similar values. The amount of the He ash required is indeed similar to the case for MUSSES1604D, as these two SNe could share similar features. \citet{cao2016} noted that another SN 2002es-like SN Ia, iPTF14dpk, does not show the early flash while otherwise looks similar to iPTF14atg. They suggested that SN 2002es-like objects may be accompanied in general by the companion interaction, while its detectability depends on the viewing direction. We note that the similar configuration would also apply to the He detonation scenario, as the He detonation may also introduce a large asymmetry in the distribution of the He detonation ash \citep{fink2010,kromer2010}. 

In sum, the natures of the early flash of iPTF14atg allow two interpretations; the companion interaction as was originally suggested by \citet{cao2015} and the He detonation. The two interpretations differ in the {\em UV} property in the earlier phase, and this is key in distinguishing the two scenarios. Another important consideration is on the properties of the maximum light, which is indeed one of the main arguments for the He detonation model for MUSSES1604D. The maximum luminosity associated with the main SN component is the main difference between iPTF14atg and MUSSES1604D. At the peak magnitude of $\sim -18$ mag in the $V$-band for iPTF14atg, we expect substantial absorption especially in the blue including the {\em UV} band passes for the He detonation scenario (Figure 11). Even Model 11A with the $V$-band peak brighter than $-19$ mag shows substantial absorption in the $UVW1$-band by $\sim 1$ mag as compared to Model 11N. Model 8A, with the peak $V$-band magnitude of $\sim -18$ mag, shows that the emission in the $UVW1$-band is totally blacked out. Therefore, for the mass of the He detonation ash of $\gsim 0.02 M_{\odot}$ required to explain the nature of the early flash in this scenario, the expectation is that the main SN component will be very faint in the {\em UV} around the maximum-light.  

We emphasize that the present study does not aim at providing a complete picture of the He detonation scenario from early to late, especially in the {\em UV} as the nature of the {\em UV} line blending is still theoretically uncertain (note that even our model sequence N, without the He detonation, is unable to explain the observed maximum-phase {\em UV} behavior).  Indeed, the maximum light behavior has not been investigated in detail in the {\em UV} even for normal SNe Ia, and our understanding is totally lacking for the {\em UV} behaviors of different sub-types of SNe Ia. Still, if MUSSES1604D is to be interpreted as a He detonation-triggered SN, then the similar amount of the He detonation ash associated with a sub-luminous SN should be accompanied by the stronger absorption around the maximum light, and this is phenomenologically not compatible to the observed properties of iPTF14atg. Therefore, the companion interaction scenario is possibly favored for iPTF14atg, while further investigations for the He detonation scenario with a range of the parameter space will be required to robustly identify the origin of the early flash found for iPTF14atg. 

\subsection{SNe 2017cbv and 2012cg}

\begin{figure*}
\begin{center}
\hspace{-2cm}
        \begin{minipage}[]{0.45\textwidth}
                \epsscale{1.6}
                \plotone{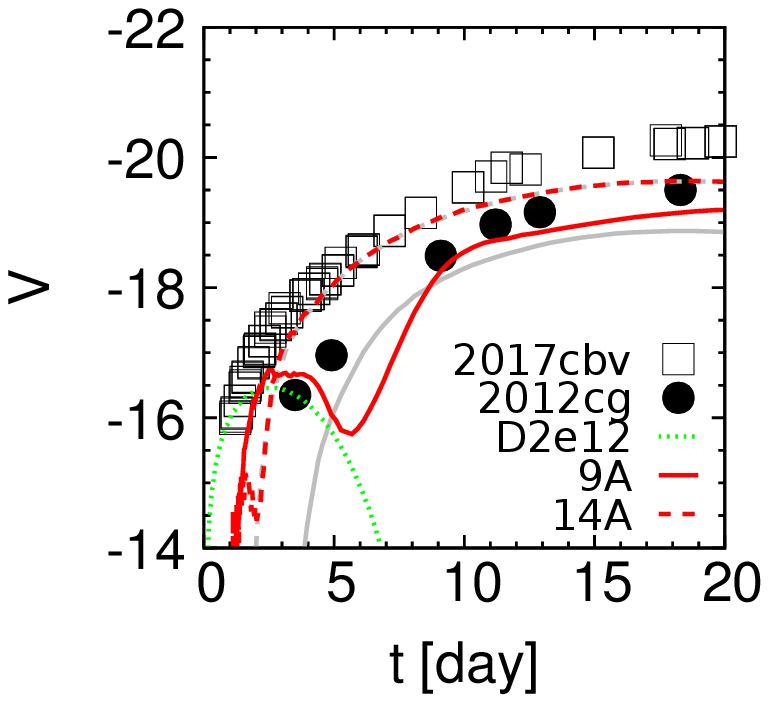}
        \end{minipage}
        \begin{minipage}[]{0.45\textwidth}
                \epsscale{1.6}
                \plotone{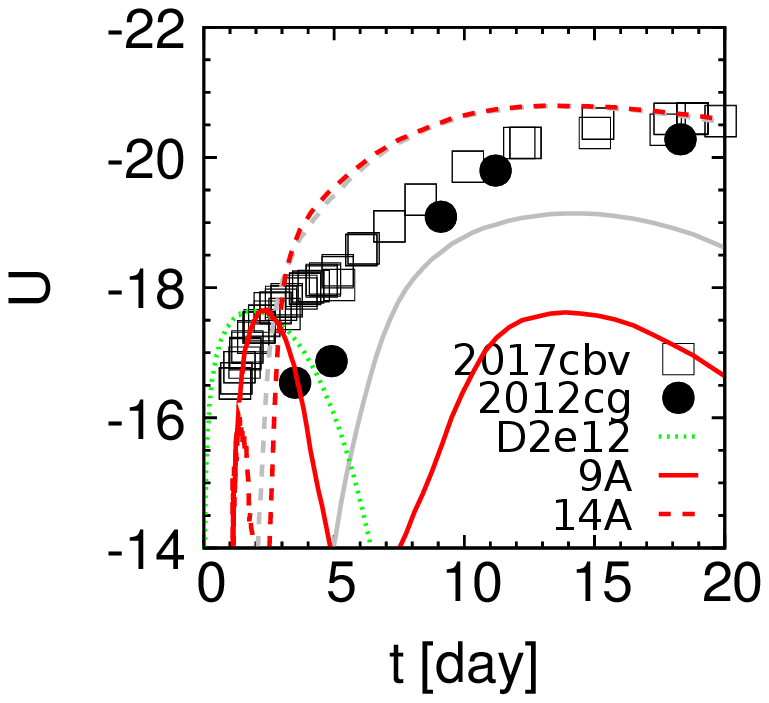}
        \end{minipage}\\
\hspace{-2cm}
        \begin{minipage}[]{0.45\textwidth}
                \epsscale{1.6}
                \plotone{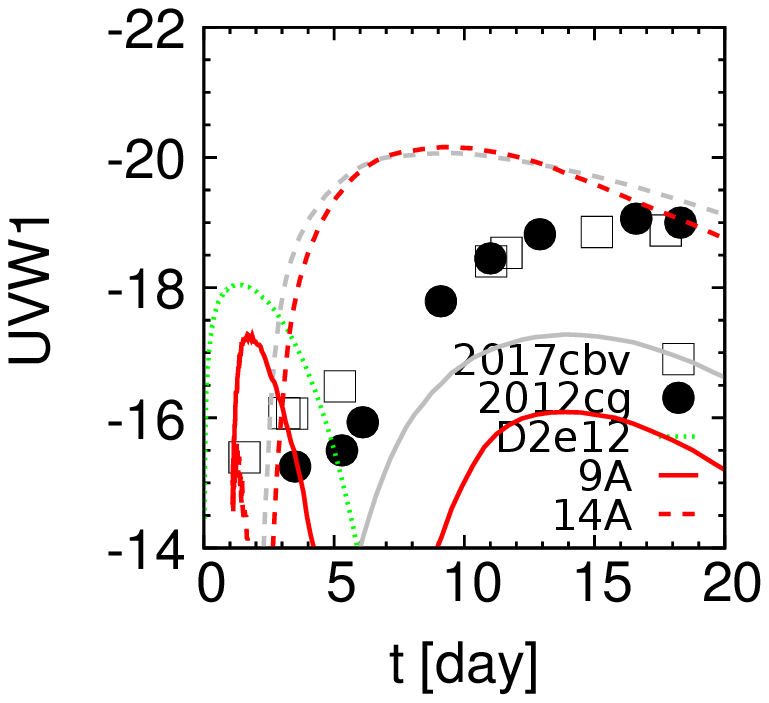}
        \end{minipage}
\end{center}
\caption
{The $V$-band, $U$-band, and $UVW1$-band light curves of SNe 2017cbv \citep[open squares;][]{hoss2017} and 2012cg \citep[filled circles;][]{marion2016}, as compared to a few selected models. Shown here are the companion interaction model (with the separation of $2 \times 10^{12}$ cm; green dotted), the He detonation models 9A (red solid) and 14A (red dashed). The corresponding models without the surface He detonation products, 9N and 14N, are also shown (gray, with the line styles following the corresponding models 9A and 14A). For the He detonation model, the explosion date is set at 1 day in this plot (i.e., the putative explosion date shifted later by 1 day from the fiducial estimate given by \citet{marion2016} and \citet{hoss2017}). 
\label{fig12}}
\end{figure*}

The $V$, $U$, and $UVW1$-band light curves of SNe 2017cbv \citep{hoss2017} and 2012cg \citep{marion2016} are shown in Figure 12. We adopt the distance moduli of 31.14 mag and 30.90 mag for SNe 2017cbv and 2012cg, respectively. The extinction corrections adopted here are $E(B-V) = 0.15$ mag and $0.2$ mag, respectively, with $R_{V}= 3.1$. Note that the former accounts only for the MW extinction, while the latter is dominated by the host contribution\footnote{The extinction is uncertain. Because of this uncertainty, \citet{hoss2017} compared these two SNe only considering the MW extinction. If we adopt $E (B-V) = 0.2$ mag for SN 2012cg including the possible host extinction \citep{marion2016}, these two SNe are relatively similar in the peak light curves (Fig. 12).}. 

For SN 2017cbv, the time scale of $\sim 5$ days for the early excessive emission is clear thanks to the very intensively sampled light curves. This is too long for the CSM interaction scenario, but can be roughly consistent with the companion interaction scenario. However, as already noted by \citet{hoss2017}, the companion interaction model predicts too bright {\em UV} emission. While the contribution of the underlying main SN component during this early phase is unknown, we could place a conservative estimate in the observed {\em UV}-optical color during the early phase. The $V$-band magnitude must have contributed at least at a level of $\sim -16$ mag, and the upper limit in the $UVW1$-band for the excessive emission is $\sim -16$ mag. Therefore, the $UVW1-V$ color must be red with $UVW1-V \gsim 0$ even conservatively, during the first 5 days. This is against the expected blue color from the companion interaction model, i.e., $UVW1-V \lsim -1$ for which the magnitude at each peak is used (Figure 6). Figure 12 shows some model examples, which show that the $UVW1$-band magnitude, or even the $U$-band magnitude, should have easily exceeded the observed limit. 

This problem was indeed already noted by \citet{hoss2017}. They speculated that possible line blanketing may suppress the {\em UV} emission, while also discussing other scenarios as summarized in the beginning of \S 7. We note that this possibility is unlikely: (1) This hypothesized absorption will need to work more effectively in the earlier phase since the expected `delay' between the {\em UV} peak to the optical peak has not been identified. (2) Related to this, the temperature decrease (or SED evolution) is much slower than the prediction by the companion (or CSM) interaction scenario, i.e., $T_{\rm r} \propto t^{-1}$ (see eq. 7) in the adiabatic loss limit. (3) The luminosities initially monotonically increase in all the bands from the {\em UV} to the optical, thus the quasi-bolometric luminosity does not seem to show the decrease predicted by the companion (or CSM) interaction. The observed behaviors thus suggest that the bolometric luminosity does {\em increase} initially, and the temperature is not decreasing rapidly. These are qualitatively more consistent with the He detonation scenario than the companion/CSM interaction scenarios. 

However, quantitatively even the sequence of the He detonation models examined in this paper predicts too bright {\em UV} emission, or equivalently too blue {\em UV}-optical color (Figure 6). This is illustrated by a few models shown in Figure 12. We selected models which result in relatively over-luminous SNe. The inspection of the model behaviors in the initial flash phase suggests that it would be remedied if the diffusion time scale would be further increased while not increasing the energy input. This would be possible in the context of the He detonation scenario, if the mass of the non-radioactive species would be increased. At the same time, adding the Fe-peak elements or Ca/Ti is not favored, as this would create too much absorption in the blue around the maximum light (see Figure 12; again note that understanding the maximum-phase light curves, especially in UV, of bright/faint SNe Ia will require further study). Such a configuration might be possible within the He detonation context for a relatively low mass WD with a thick He layer, which would leave He as a dominant species. This would however contradict to the standard double detonation model sequence, as an over-luminous SN Ia like SN 2017cbv (but note that there is an uncertainty in the distance and thus the luminosity) is expected to be an outcome of a relatively massive WD with at least $\sim 1 M_{\odot}$ (see Tab. 2). 

Alternatively, a similar configuration might be more naturally expected by a mixing of $^{56}$Ni outward even without the He detonation. It is indeed indicative that SN 2017cbv is potentially an over-luminous SN Ia (i.e., SN 1991T/1999aa-like) while it shares similar spectra with normal SNe Ia. Abundance tomography of SN 1991T-like objects generally requires an extended distribution of $^{56}$Ni out to the surface \citep{sasdelli2014,zhang2016}. 
The non-monotonic light curve will require non-monotonic distribution of $^{56}$Ni \citep[e.g.,][]{piro2013,piro2014} which favors a large scale asymmetry \citep[e.g.,][]{maeda2010} rather than a small scale clumping. A combined analysis of the light curve and spectral sequence analysis will be required to clarify further the origin of the early flash. 

Similar arguments apply to SN 2012cg. If the extinction of $E(B-V) \sim 0.2$ mag is adopted, the light curves of SN 2012cg are similar to SN 2017cbv around the maximum. The early flash is likely fainter by $\sim 0.5 - 1$ mag in all the bands, but the color at this phase is similar. The light curve data obtained for SN 2012cg alone would not readily reject either of the companion interaction scenario or the He detonation (or $^{56}$Ni mixing) scenario, both of which can roughly explain the features of the first light curve points. This highlights the importance of the intensively sampled light curves, like SN 2017cbv, to clarify the nature of the early flash (or early excessive emission). 

Features in the very early-phase spectra of SN 2012cg \citep{marion2016} may also favor the models which involve the radioactive heating rather than the shock thermalization. Its spectrum at $\sim -16$ days measured from the $B$-band maximum shows a blue spectrum plus relatively shallow absorption features. The absorption features are, however, not as shallow as expected from a simple sum of a blue BB continuum plus a template normal SN spectrum as already noted by \citet{marion2016}. The development of the absorption lines for the initial flash is expected for the He detonation scenario (\S 5), and this behavior could also be shared by the scenario involving the extensive $^{56}$Ni mixing. 

We note that similar arguments would also apply to another example, iPTF16abc, which shows blue color in the optical in the initial phase (before $\sim -13$ days measured from the $B$-band maximum) \citep{miller2017}. It however shows much redder {\em UV} - optical color than iPTF14atg, sharing the similar behavior with SNe 2012cg and 2017cbv. This led \citet{miller2017} to conclude that the most likely explanation is either the extensive $^{56}$Ni mixing or interaction with diffuse (extended) CSM. Our most favored scenario is the $^{56}$Ni mixing, following the same arguments for SNe 2012cg and 2017cbv while the He detonation scenario may not be robustly rejected. For the case of iPTF16abc, the short-lived C II absorption may support the extensive $^{56}$Ni mixing \citep{miller2017}. The difference in iPTF16abc and SNe 2012cg/2017cbv may then be attributed to the different degree of the $^{56}$Ni mixing, where a larger amount of the unburnt C+O WD materials near the surface may be expected for a less luminous SN Ia \citep[e.g.,][]{zhao2016}. 

\section{Conclusions}
In this paper, we have investigated different scenarios to create excessive `early-phase' emission from SNe Ia within the first few days after the explosion. We have first provided simple prescriptions to describe main physical processes involved in the interaction with either CSM \citep{piro2016} or a companion star \citep{kasen2010} \citep[but see][]{kutsuna2015}. We have shown that these two scenarios can be understood in the same context. In the CSM interaction scenario, the main function is the mass of the CSM (which determines both the time scale and the initial thermal energy content), while in the companion interaction model it is the binary separation (which determines the initial thermal energy density). In general, these scenarios behave in the same way with the `cooling emission' from CCSNe, which has been extensively studied by several groups previously \citep[e.g.,][]{arnett1980,bersten2012}, and it is characterized by the decrease both in the bolometric luminosity and radiation temperature. 

The He detonation scenario has been recently suggested as an alternative scenario to create the early excessive emission \citep{jiang2017,noebauer2017}. The key physical process involved in this scenario is different than the CSM/companion interaction scenarios. The He detonation scenario predicts that the bolometric luminosity initially increases and peaks in the first few days. The evolution in the color is generally slower than the other scenarios, and the color tends to be redder. While the relation between the power source and opacity source is different than in the interaction scenarios, the He detonation scenario results in the emission whose peak properties, for a given peak date, can be qualitatively similar to those expected for the interaction scenarios. Details including the temporal evolution (e.g., the $B-V$ color evolution) can be different, depending on the model parameters involved in the He detonation scenario. 

The difference is highlighted in the {\em UV} behaviors. The He detonation scenario predicts that the {\em UV} peak is reached substantially earlier than the companion/CSM interaction scenarios, typically within one day after the explosion for a range of the model parameters. The {\em UV} emission at the time of the associated optical peak is weaker than the interaction scenarios, with the difference especially evident by comparing different models creating the similar optical peak dates. Accordingly, the {\em UV}-optical color is red. Especially useful information can be obtained by the {\em UV}-optical color evolution, as this observable reflects the main difference in the underlying physical processes and it is free from the uncertainty in the extinction.

Spectra in the early flash phase for the He detonation scenario have been presented. At the peak of the optical emission, the {\em UV} line absorptions already start developing. Further development of the absorption lines from the He detonation ash (Fe-peaks, Ca, Ti) is seen in the decay phase after the initial peak, the strength of which depends not only the mass of the He detonation products but also the luminosity of the underlying main SN ejecta component. 

Around the maximum phase, the presence of the He detonation products in the outermost layer of the SN ejecta can change the SN appearance. For a larger amount of the He detonation products and/or the lower luminosity from the main SN ejecta component, the effect becomes larger. Within the `minimal He mass' model sequence within the double detonation scenario \citep[i.e., a larger amount of the He layer for a less massive WD, therefore for lower peak luminosity;][]{fink2010,woosley2011}, this effect for the fainter SNe (i.e., the les massive WD in the adopted double detonation sequence) will essentially black out the {\em UV} and the blue portion of the optical spectra, resulting in an extremely red SN. For relatively brighter SNe (e.g., with $M$($^{56}$Ni) $\gsim 0.6 M_\odot$, which is related to the He mass of $\lsim 0.05 M_\odot$ in our model sequence), the effect is not strong at the maximum brightness, and the effect of the He detonation products may indeed be hidden. Still, the absorptions tend to develop in the post-maximum decay phase (except for the most massive WD progenitors), which could be used as additional diagnostics of the He detonation scenario. These features suggest that (1) we generally expect a relation between the properties of the early-phase flash and those of the maximum SN emission, in a way the brighter and slower initial flash is accompanied by a more substantial effect of the additional absorptions (and reddening) by the He detonation products, (2) this relation, however, should be considered together with the maximum luminosity of the SN, since larger luminosity suppresses the effect of the additional absorption, and that (3) the effect is especially pronounced in the {\em UV} both in the early and maximum phases. 

We have further discussed the origin of the early excessive emission reported for a few SNe so far. The properties of MUSSES1604D \citep{jiang2017} fit into the model predictions by the He detonation scenario. The properties of the first example of the reported early-phase excess, iPTF14atg \citep{cao2015,cao2016}, are consistent either with the companion interaction or the He detonation scenario. In the He detonation scenario, it requires the mass of the He layer $\gsim 0.02 M_{\odot}$, similar to the estimate for MUSSES1604D. This would, however, result in a very red maximum emission due to the absorption by the He detonation products, given the sub-luminous nature of iPTF14atg, being possibly inconsistent with the observed property. While the detailed modeling of individual SNe, especially in the maximum phase, is beyond the scope of this paper, this consideration raises a doubt on the He detonation scenario for iPTF14atg. 

SNe 2012cg \citep{marion2016} and 2017cbv \citep{hoss2017}, while the spectra show some differences, seem to share similar light curve evolution around the maximum light, potentially belonging to a luminous class like SN 1991T. The early-phase excessive emission is stronger in SN 2017cbv than SN 2012cg. For SN 2012cg, it is possible to explain the data either by the companion interaction or the He detonation scenario. Further investigation, however, is limited by the sparse data set to characterize the nature of the early emission. Indeed, the densely sampled light curves of SN 2017cbv indicate that the {\em UV} emission is much weaker than that expected by the interaction scenario(s). Redistribution of the radiation energy by the {\em UV}-line absorptions within the interaction scenario is unlikely, since we do not see the characteristic property of the interaction scenario, i.e., decreasing bolometric luminosity. Rather, it is more likely explained by a scenario involving initially increasing bolometric luminosity, as in the He detonation scenario. One would construct the He detonation model to explain these behaviors by considering a thick He layer dominated by non-radioactive species beyond our standard `double detonation' model sequence. A more natural and favored origin of the excessive early-phase emission found in these SNe is an extensive mixing of $^{56}$Ni in the main SN ejecta, which would be consistent with the nature of these SNe as (possibly) over-luminous SNe. Another example showing a similar early-phase behavior, iPTF16abc \citep[with normal luminosity;][]{miller2017}, probably shares the same mechanism with SNe 2012cg and 2017cbv for the early emission. The detection of C II in the very early-phase spectra might support this hypothesis (e.g., a less extended distribution of $^{56}$Ni for a fainter object). 

For the companion interaction scenario, we point out a possible contribution to the early flash by the inner $^{56}$Ni exposed to the observer direction. Qualitatively, this effect may well become significant for the small binary separation. This effect will increase the optical luminosity in the early flash while not strongly affect the {\em UV} behavior. The color will still be bluer than the He detonation scenario. For MUSSES1604D, this effect was indeed included in the simulation shown by \citet{jiang2017} and it was concluded that the companion interaction scenario is inconsistent with the observed data for this SN. For iPTF14atg, our companion interaction model presented in this paper without this effect provides reasonable explanation of the observed light curves. Indeed, the expected luminosity from the inner $^{56}$Ni may compete to the thermal energy generation, and therefore the self-consistent model including both contributions will be necessary for detailed modeling of this object. For SNe 2012cg and 2017cbv (and iPTF16abc), this additional effect alone will not fully remedy the difficulty in the companion interaction scenario to explain these SNe; even the He detonation scenario (which will lead to the larger optical contribution than this scenario for a given {\em UV} luminosity) predicts too large {\em UV} luminosity as compared to the optical. 

\acknowledgements 
The work has been supported by Japan Society for the Promotion of Science (JSPS) KAKENHI Grant 18H04585 and 17H02864 (K.M.), 16H01087 and 26287029 (J.J. and M.D.),  18J12714 (J.J.). 16H06341, 16K05287, and 15H02082 (T.S.). Simulations for the He detonation models were carried out on a Cray XC30 at the Center for Computational Astrophysics, National Astronomical Observatory of Japan. 

\appendix
\section{Radioactive Decay Energy Input}
$^{52}$Fe decays into the metastable $^{52m}$Mn either by the electron capture or the positron emission (with the branching ratio of 55\% and the average kinetic energy of 340 keV), followed mainly by a 168.7 keV line emission. The e-folding time ($\tau_{52Fe}$) is $0.498$ day. $^{52m}$Mn then decays with $\tau_{52Mn} = 0.021$ day through positron emission with the average kinetic energy of 1.174 Mev followed by a 1.434 MeV line emission. Given the short life time of the daughter nucleus, the decay chain can be treated as a single step. Including the 511 keV annihilation photons and other minor $\gamma$-ray lines (as included in our simulations), the (bolometric) energy input from the decay chain $^{52}$Fe $\to$ $^{52}$Mn $\to$ $^{52}$Cr can be described as follows: 
\begin{eqnarray}
L(^{52}{\rm Fe/Mn/Cr}) & = & \left(s_{\gamma, {\rm Fe/Mn}} +s_{{\rm e+}, {\rm Fe/Mn}}\right) e^{-t/\tau_{52Fe}} \ , \\ 
s_{\gamma, {\rm Fe/Mn}} & = & 2.7 \times 10^{45} {\rm erg \ s}^{-1} \left(\frac{M(^{52}{\rm Fe})}{M_{\odot}}\right) \ , \\
s_{{\rm e+}, {\rm Fe/Mn}} & = & 1.1 \times 10^{45} {\rm erg \ s}^{-1} \left(\frac{M(^{52}{\rm Fe})}{M_{\odot}}\right) \ , 
\end{eqnarray}
where $M$($^{52}$Fe) is the initial mass of $^{52}$Fe before the decay. 

$^{48}$Cr decays into $^{48}$V through the electron capture with $\tau_{48Cr} = 1.30$ day,  followed by 112 keV and 308 keV lines. $^{48}$V then decays into  $^{48}$Ti with $\tau_{48V} = 23.0$ day, either by the electron capture or the positron emission ($50\%$). The average positron kinetic energy is 290 keV. This is then followed mainly by 984 keV and 1312 keV line emissions. Including the 511 keV annihilation photons and other minor $\gamma$-ray lines, the (bolometric) energy input from the decay chain $^{48}$Cr $\to$ $^{48}$V $\to$ $^{48}$Ti can be described as follows: 
\begin{eqnarray}
L(^{48}{\rm Cr/V/Ti}) & = & s_{\gamma, {\rm Cr}} e^{-t/\tau_{48Cr}} + (s_{\gamma, {\rm V}} + s_{{\rm e+}, {\rm V}}) (e^{-t/\tau_{48V}} - e^{-t/\tau_{^{48}{\rm Cr}}}) , \\ 
s_{\gamma, {\rm Cr}} & = & 1.5 \times 10^{44} {\rm erg \ s}^{-1} \left(\frac{M(^{48}{\rm Cr})}{M_{\odot}}\right) \ , \\
s_{\gamma, {\rm V}} & = & 6.2 \times 10^{43} {\rm erg \ s}^{-1} \left(\frac{M(^{48}{\rm Cr})}{M_{\odot}}\right) \ , \\
s_{{\rm e+}, {\rm V}} & = & 3.1 \times 10^{42} {\rm erg \ s}^{-1} \left(\frac{M(^{48}{\rm Cr})}{M_{\odot}}\right) \ ,
\end{eqnarray}
where $M$($^{48}$Cr) is the initial mass of $^{48}$Cr before the decay.

\end{document}